\documentstyle[aps,eqsecnum,epsf,epsfig,prd,preprint]{revtex}
\newcommand{\nd}[1]{/\hspace{-0.6em} #1}
\begin{document}

\baselineskip=18pt

\preprint{\baselineskip=12pt{\vbox{\hbox{OUTP-99-47P}
\hbox{CERN-TH/99-256} \hbox{NBI-HE-99-34}
\hbox{hep-th/9909129} \hbox{~~~} \hbox{September 1999}}}}
\title{Non-linear Schr\"odinger Dynamics of Matrix D-branes}
\author{Nick E. Mavromatos}
\address{\vspace{2mm}\baselineskip=12pt Department of Physics -- Theoretical
Physics, King's College London\\The Strand, London WC2R 2LS, U.K.\\
{\tt nikolaos.mavromatos@cern.ch}}
\author{Richard J. Szabo}
\address{\vspace{2mm}\baselineskip=12pt The Niels Bohr Institute\\ Blegdamsvej
17, DK-2100 Copenhagen \O, Denmark\\ {\tt szabo@nbi.dk}}
\maketitle
\begin{abstract}
\baselineskip=12pt

We formulate an effective Schr\"odinger wave equation describing the quantum
dynamics of a system of D0-branes by applying the Wilson renormalization group
equation to the worldsheet partition function of a deformed $\sigma$-model
describing the system, which includes the quantum recoil due to the exchange of
string states between the individual D-particles. We arrive at an effective
Fokker-Planck equation for the probability density with diffusion coefficient
determined by the total kinetic energy of the recoiling system.
We use Galilean invariance of the system
to show that there are three possible solutions of the associated non-linear
Schr\"odinger equation depending
on the strength of the open string interactions among the D-particles. When
the open string energies are small compared to the total kinetic energy of the
system, the solutions are governed by freely-propagating solitary waves. When
the string coupling constant reaches a dynamically determined critical value,
the system is described by minimal uncertainty wavepackets which describe the
smearing of the D-particle coordinates due to the distortion of the surrounding
spacetime from the string interactions. For strong string interactions, bound
state solutions exist with effective mass determined by an energy-dependent
shift of the static BPS mass of the D0-branes.

\end{abstract}

\vfill
\pagebreak

\font\mathsm=cmmi9
\font\mybb=msbm10 at 12pt
\def\bb#1{\hbox{\mybb#1}}
\font\mybbs=msbm10 at 9pt
\def\bbs#1{\hbox{\mybbs#1}}
\def\e{{\rm e}}

\newcommand{\complex}{{\bb C}} 
\newcommand{\complexs}{{\bbs C}} 
\newcommand{\zed}{{\bb Z}} 
\newcommand{\real}{{\bb R}} 
\newcommand{\reals}{{\bbs R}} 
\newcommand{\zeds}{{\bbs Z}} 
\newcommand{\mod}{{\bb M}} 
\newcommand{\mods}{{\bbs M}} 
\newcommand{\be}{\begin{equation}}
\newcommand{\ee}{\end{equation}}
\newcommand{\bea}{\begin{eqnarray}}
\newcommand{\eea}{\end{eqnarray}}
\newcommand{\ra}{\rightarrow}
\newcommand{\tr}{\mbox{tr}}
\newcommand{\Tr}{\mbox{Tr}}
\newcommand{\al}{\alpha}
\newcommand{\bt}{\beta}
\newcommand{\bz}{{\bar{z}}}
\newcommand{\del}{\Delta}
\newcommand{\Th}{\Theta}
\newcommand{\td}{\tilde{\del}}
\newcommand{\g}{\gamma}
\newcommand{\bchi}{\bbox{\chi}}
\newcommand{\nn}{\nonumber}
\newcommand{\NO}{\,\mbox{$\circ\atop\circ$}\,} 
\newcommand{\semiplus}{{\supset\!\!\!\!\!\!\!+~}} 

\section{Introduction and Summary}

Despite the enormous amount of activity over the past few years towards
understanding the dynamics of Dirichlet $p$-branes \cite{polchinski}, the
problem of demonstrating that a system of $N$
moving D-particles can form a bound
state is still unresolved. It is relevant to the Matrix Theory conjecture
\cite{bfss} in which these particles are interpreted as Kaluza-Klein modes of
11-dimensional M-Theory compactified on a circle and are described by the
supersymmetric $N\times N$ matrix quantum mechanics that is obtained from
dimensional reduction of ${\cal N}=1$ supersymmetric Yang-Mills theory in ten
dimensions. The existence of such a tower of states is equivalent to the
statement that the quantum mechanics admits exactly one bound state for each
$N$. The original threshold bound state problem was addressed in \cite{kpdfs}
and recent progress has been made in \cite{bound}. However, beyond the proof of
existence in the case $N=2$, the general case $N>2$ remains in large part an
open problem. In this paper we shall address the bound state problem for a
system of $N$ non-relativistic, recoiling D0-branes by studying their moduli
space dynamics in certain limits.

The effective worldvolume dynamics of a single D$p$-brane coupled to a
worldvolume gauge field and to background supergravity fields is described by
the action \cite{polchinski}
\be
I_{{\rm D}p}={\cal T}_p\int
d^{p+1}\sigma~\e^{-\phi}\,\sqrt{-\det_{\alpha,\beta}\left[G_{\alpha\beta}
+B_{\alpha\beta}+2\pi\alpha'F_{\alpha\beta}\right]}+{\cal T}_p\int
d^{p+1}\sigma~\left[C\wedge\e^{2\pi\alpha'F+B}\wedge{\cal G}\right]_{p+1}
\label{IDp}\ee
The first term in (\ref{IDp}) is the Dirac-Born-Infeld action with ${\cal T}_p$
the $p$-brane tension, $\alpha'$ the string Regge slope, $\phi$ the dilaton
field, $F=dA$ the worldvolume field strength, and $G$ and $B$ the pull-backs of
the target space metric and Neveu-Schwarz two-form fields, respectively, to the
D$p$-brane worldvolume. It is a generalization of the geometric volume of the
brane trajectory. The second term is the Wess-Zumino action (restricted to its
$p+1$-form component) with $C$ the pullback of the sum over all electric and
magnetic Ramond-Ramond (RR) form potentials and $\cal G$ a geometrical factor
accounting for the possible non-trivial curvature of the tangent and normal
bundles to the $p$-brane worldvolume. It describes the coupling of the
D$p$-brane to the supergravity RR $p+1$-form fields as well as to the
topological charge of the worldvolume gauge field and to the worldvolume
gravitational connections. The fermionic completion of
the action (\ref{IDp}), compatible with spacetime supersymmetry and worldvolume
$\kappa$-symmetry, has been described in \cite{susybi}. For a recent review of
the Born-Infeld action and its various extensions in superstring theory, see
\cite{birev}.

While the generalization of the Wess-Zumino Lagrangian to multiple D$p$-branes
is obvious (one simply traces over the worldvolume gauge group in the
fundamental representation), the complete form of the non-abelian Born-Infeld
action is not known. In \cite{tseytlin} it was proposed that the background
independent terms can be derived using $T$-duality from a 9-brane action
obtained from the corresponding abelian version by symmetrizing all gauge
group traces in the vector representation \cite{birev}. A
direct calculation of the leading terms in a weak supergravity background has
been calculated using Matrix Theory methods in \cite{van}. Based on the Type I
formulation, i.e. by viewing a D-particle in the Neumann picture and imposing
$T$-duality as a functional canonical transformation in the string path
integral \cite{dorn}, the effective moduli space Lagrangian was derived in
\cite{ms1} and shown to coincide (to leading orders in a velocity expansion)
with the non-abelian Born-Infeld action of \cite{tseytlin}. In the following we
will use this moduli space approach to D-brane dynamics to describe some
properties of the multiple D-brane wavefunction.

The novel aspect of the approach of \cite{ms1} is that the moduli space
dynamics induces an effective target space geometry for the D-branes which
contains information about the short-distance spacetime structure probed by
multiple D-particles. Based on this feature, string-modified spacetime and
phase space uncertainty relations can be derived and thereby represent a proper
quantization of the noncommutative spacetime seen by low-energy D-particle
probes \cite{ms2}. The crucial property of the derivation is the incorporation
of proper recoil operators for the D-branes and the short open string
excitations connecting them. The smearing of the spacetime coordinates $y_i^a$
(in general $i=1,\dots,9-p$ label the transverse coordinates of the D$p$-brane
and $a=1,\dots,N$ the component branes of the multiple D-brane configuration)
of a given D-particle as a result from its open string interactions with other
branes can be seen directly from the formula for the variance
\be
\left(\Delta
y_i^a\right)^2\equiv\left[\left(Y_i-Y_i^{aa}\,I_N\right)^2\right]^{aa}
=\sum_{b\neq a}|Y_i^{ab}|^2
\label{Deltayia}\ee
where $Y_i^{ab}$ are the $u(N)$-valued positions of the D-particles ($a=b$) and
of the open strings connecting branes $a$ and $b$ ($a\neq b$), and $I_N$ is the
$N\times N$ identity matrix. The recoil operators give a relevant deformation
of the conformal field theory describing free open strings, and thus lead to
non-trivial renormalization group flows on the moduli space of coupling
constants. The moduli space dynamics is thereby governed by the Zamolodchikov
metric and the associated $C$-theorem. Physically, the recoil operators
describe the appropriate change of quantum state of the D-brane background
after the emission or absorption of open or closed strings. They are a
necessary ingredient in the description of multiple D-brane dynamics, in which
coincident branes interact with each other via the exchange of open string
states. The quantum uncertainties derived in \cite{ms1,ms2} were found to
exhibit quantum decoherence effects through their dependence on the recoil
energies of the system of D-particles. This suggests that the appropriate
quantum dynamics of D0-branes should be described by some sort of stochastic
string field theory involving a Fokker-Planck Hamiltonian.

As in \cite{birev,tseytlin}, the derivation in \cite{ms1} assumes constant
background supergravity fields. However, another important ingredient missing
in the moduli space description are the appropriate residual fermionic terms
from the supersymmetry of the initial static D-brane configuration. While the
recoil of the D-branes breaks supersymmetry, it is necessary to include these
terms to have a complete description of the stability of the D-particle bound
state. As shown in \cite{kallosh}, the energy of the bound states of D-branes
and strings is determined by the central charge of the corresponding spacetime
supersymmetry algebra. Nonetheless, the bosonic formalism that we display below
can be exploited to a large extent to describe at least heuristically the
quantum phase structure of the multiple D-particle system and, in particular,
determine the mass and stability conditions of the candidate bound state. One
reason that this approach is expected to yield reliable results is that we view
the system of D-branes and strings as a quantum mechanical system (rather than
a quantum field theoretical system as might be the case from the fact that
$T$-duality is used to effectively integrate over the transverse coordinates of
the branes), with the D-brane recoil constituting an excitation of this system.
The recoiling system of D-branes and strings can be viewed as an excited state
of a supersymmetric (static) vacuum configuration.
The breaking of target space supersymmetry by the excited state of the
system may thereby constitute a symmetry obstruction situation in the spirit
of \cite{symob}. According to the symmetry
obstruction hypothesis, the ground state of a system of
(static) strings and D-branes is a BPS state,
but the excited (recoiling) states do not respect the supersymmetry
due to quantum diffusion and other effects.
Phenomenologically, the supersymmetry
breaking induced by the excited system of recoiling D-particles
will distort the spacetime surrounding them
and may result in a decohering spacetime foam,
on which low energy (point-like) excitations live. This motivates the study
of non-supersymmetric D-branes recoiling under the exchange of strings.
Such quantum mechanical systems exhibit diffusion
and may be viewed as non-equilibrium (open) quantum systems,
with the non-equilibrium state being related naturally to the picture
of viewing the recoiling D-brane system as an excited state of some
(non-perturbative) supersymmetric D-brane vacuum configuration.

The main relationship we shall exploit in obtaining the quantum dynamics of
multiple D-particle systems is that between the Dirichlet
partition function in
the background of Type II string fields and the semi-classical (Euclidean)
wavefunctional $\Psi[Y^i]$ of a D$p$-brane.
This relation is usually expressed as \cite{hlp6,emn2}
\be
{\cal Z}=\int DY^i~\Psi[Y^i]
\label{Dppartfn}\ee
The wavefunction $\Psi[Y^i]$ is expressed in terms of the generating functional
which sums up all one-particle irreducible connected worldsheet diagrams whose
boundaries are mapped onto the D-brane worldvolume. Integration over the
worldvolume gauge field is implicit in $\Psi$ to ensure Type II winding number
conservation. Dirichlet string perturbation theory yields
\be
\Psi[Y^i]=\exp\sum_{h=1}^\infty\e^{(h-2)\phi}\,{\cal S}_h[Y^i]
\label{Dpert}\ee
where ${\cal S}_h$ denotes the amplitude with $h$ holes, in which an implicit
sum over handles is assumed. However, as we will discuss in the following,
the identification (\ref{Dppartfn}) is {\it not} the only one
consistent with the approach to D-brane dynamics advocated in
\cite{ms1}, and one may instead identify the worldsheet Dirichlet partition
function, summed over all genera, with the probability
distribution corresponding to the wavefunction $\Psi$.
Using this identification, the Wilson
renormalization group equation has been proposed as a defining principle for
obtaining string field equations of motion, including the appropriate
Fischler-Susskind mechanism for the contributions from higher genera
\cite{hlp6}. When applied to Dirichlet string theory, we shall find that the
consistent D-brane equation of motion follows from the renormalization group
equation.

More precisely, within the framework of a perturbative logarithmic conformal
field theory approach to multiple D-brane dynamics \cite{ms1}, we will show
that the intricate quantum dynamics of a system of interacting D-particles is
described by a non-linear Schr\"odinger wave equation. The corresponding
probability density is of the Fokker-Planck type, with quantum diffusion
coefficient $\cal D$ given by the square of the modulus of the recoil velocity
matrix of the bound state system of D-particles and strings:
\be
{\cal D}=c_G\,\sqrt{\alpha'}\,\sum_{i=1}^9\tr\,|\bar U^i|^2
\label{diffintro}\ee
where $c_G$ is a numerical constant and
$\bar U_{ab}^i$ is the (renormalized) constant velocity matrix of a system of
$N$ D-particles arising due to the D-particle recoil from the scattering of
string states. This phenomenon is in fact characteristic of Liouville string
theory, on which the above approach is based. Since the D-particle interactions
distort their surrounding spacetime, these non-linear structures may be thought
of as describing short-distance quantum gravitational properties of the D-brane
spacetime. Non-linear equations of motion for string field theories have been
derived in other contexts in \cite{cst}. From this nonlinear Schr\"odinger
dynamics we shall describe a multitude of classes of solutions, using Galilean
invariance of the D-brane dynamics which is a consequence of the corresponding
logarithmic conformal algebra. We will show that bound state solutions do
indeed exist for string couplings $g_s$ larger than a dynamically determined
critical value. The effective bound state mass is likewise determined as an
energetically induced shift of the static, BPS mass of the D0-branes.
In fact, we
shall find that there are essentially three different phases of the quantum
dynamics in string coupling constant space. Below the critical string coupling
the multiple D-brane wavefunction is described by solitary waves, in agreement
with the description of free D-branes as string theoretic solitons, while at
the critical coupling the quantum dynamics is described by coherent Gaussian
wavepackets which determine the appropriate quantum smearing of the multiple
D-particle spacetime. These results are shown to be in agreement with the
previous results concerning the structure of quantum spacetime \cite{ms1,ms2}.

We close this section by summarizing some of the generic guidelines that we
shall use in this paper for constructing a wavefunctional for the system of
D-branes. We will use a field theoretic approach by identifying the
Hartle-Hawking
wavefunction
\be
\Psi_0\simeq\e^{-S_E}
\label{hhwaveintro}\ee
where $S_E$ is the effective Euclidean action. We shall discuss the extension
to string theory and highlight the advantages and disadvantages of using this
identification. We shall also identify the probability density with the genus
expansion of an appropriate worldsheet $\sigma$-model:
\be
{\cal P}=\Psi_0^\dagger\Psi_0=\sum_{\rm genera}\int Dx~\e^{-S_\sigma[x]}
\label{probintro}\ee
The arguments in favour of this identification will be reality, and the
occurence of statistical probability distribution factors which appear in the
wormhole parameters after resummation of (\ref{probintro}) over pinched genera.
The Wilson-Polchinski worldsheet renormalization group flow, coming from the
sum over genera as in (\ref{probintro}), yields a Fokker-Planck diffusion
equation
\be
\partial_t{\cal P}={\cal D}\,\nabla^2{\cal P}-\nabla\cdot{\cal J}
\label{FPintro}\ee
where $\cal D$ is the diffusion operator defined in (\ref{diffintro}) in terms
of (renormalized) recoil velocity matrices, and $\cal J$ is the associated
probability current density. The equation (\ref{FPintro}) will follow from the
gradient flow property of the $\sigma$-model $\beta$-functions, which is also
necessary for the Helmholtz conditons or equivalently for canonical
quantization of the string moduli space.

The knowledge of the Fokker-Planck equation (\ref{FPintro}) alone does {\it
not} lead to an unambiguous construction of the wavefunction $\Psi$. There are
ambiguities associated with non-linear $\Psi$-dependent phase transformations
of the wavefunction:
\bea
\Psi&\mapsto&\e^{i{\cal N}_{\gamma,\lambda}(\Psi)}\,\Psi\nonumber\\{\cal
N}_{\gamma,\lambda}(\Psi)&=&\gamma\log|\Psi|+\lambda\,{\rm
arg}\,\Psi+\theta\Bigl(\{Y_i^{ab}\},t\Bigr)
\label{phaseintro}\eea
where $t$ is the Liouville zero mode.
Furthermore, $\Psi$ is then necessarily determined by a non-linear wave
equation if a diffusion coefficient $\cal D$ is present, as will be the case in
what follows. The non-linear Schr\"odinger equation has the form
\be
i\hbar\,\partial_t\Psi={\cal H}_0\Psi+\frac{i\hbar}2\,{\cal D}\,
\frac{\nabla^2{\cal P}}{\cal P}\,\Psi
\label{nlseintro}\ee
where ${\cal P}=\Psi^\dagger\Psi$ is the probability density. This is a
Galilean-invariant but time-reversal violating equation, exactly as expected
from previous considerations of non-relativistic D-brane dynamics and Liouville
string theory. Eq. (\ref{nlseintro}) will be the proposal in the following for
the non-linear quantum dynamics of matrix D-branes (this was noted in passing
in \cite{emn10}).

\section{Quantum Mechanics on Moduli Space}

In \cite{ms1} it was shown how a description of non-abelian D-particle
dynamics, based on canonical quantization of a $\sigma$-model moduli space
induced by the worldsheet genus expansion (i.e. the quantum string theory),
yields quantum fluctuations of the string soliton collective coordinates and
hence a microscopic derivation of spacetime uncertainty relations, as seen by
short distance D-particle probes. In the following we will proceed to construct
a wavefunction for the system of D0-branes which encodes the pertinent quantum
dynamics. To start, in this section we shall clarify certain facts about
wavefunctionals in non-critical string theories in general, completing the
discussion put forward in \cite{emn2}.

\subsection{Liouville-dressed Renormalization Group Flows}

Consider quite generally a non-critical string $\sigma$-model, defined as a
deformation of a conformal field theory $S_*$ with coupling constants
$\{g^I\}$. The worldsheet action is
\be
S_\sigma[x;\{g^I\}]=S_*[x]+\int\limits_\Sigma d^2z~g^IV_I[x]
\label{sigma1}\ee
where $V_I$ are the deformation vertex operators and an implicit sum over
repeated upper and lower indices is always understood. We assume that the
deformation is relevant, so that the worldsheet theory must be dressed by
two-dimensional quantum gravity in order to restore conformal invariance in the
quantum string theory. The corresponding Liouville-dressed renormalized
couplings $\{\lambda^I\}$ satisfy the renormalization group equations
\be
\ddot\lambda^I+Q\dot\lambda^I=-\beta^I(\lambda)
\label{beta2}\ee
where the dots denote differentiation with respect to the worldsheet zero mode
of the Liouville field. Here $Q$ is the square root of the running central
charge
deficit on moduli space and
\be
\beta^I(\lambda)=h^I\lambda^I+c^I_{~JK}\lambda^J\lambda^K+\dots
\label{betadef3}\ee
are the flat worldsheet $\beta$-functions, expressed in terms of
Liouville-dressed coupling constants. In (\ref{betadef3}), $h^I$ are the
conformal dimensions and $c^I_{~JK}$ the operator product expansion
coefficients of the vertex operators $V_I$. The minus sign in (\ref{beta2})
owes to the fact that we confine our attention here to the case of central
charge $c>25$ (corresponding to supercritical bosonic or fermionic strings).

Upon interpreting the Liouville zero mode as the target space time evolution
parameter, eq. (\ref{beta2}) is reminescent of the equation of motion for the
inflaton field $\phi$ in inflationary cosmological models \cite{l5,lm3}. In the
present case of course one has a collection of fields $\{g^I\}$, but the
analogy is nevertheless precise. The role of the Hubble constant $H$ is played
by the central charge deficit $Q$. The precise correspondence actually follows
from the gradient flow property of the string $\sigma$-model $\beta$-functions
for flat worldsheets:
\be
\beta^I=G^{IJ}\,\frac\partial{\partial g^J}C
\label{gflow}\ee
where $C=Q^2$ is the Zamolodchikov $C$-function which is associated with the
generating functional for one-particle irreducible correlation functions
\cite{mm3b}, and $G^{IJ}$ is the matrix inverse of the Zamolodchikov metric
\be
G_{IJ}=2|z|^4\left\langle V_I(z,\bar z)\,V_J(0,0)\right\rangle
\label{Zammetric}\ee
on the moduli space $\mod(\{g^I\})$ of $\sigma$-model couplings $\{g^I\}$. Then
the right-hand side of (\ref{beta2}) also corresponds to the gradient of the
potential $V$ in inflationary models:
\be
\ddot\phi+3H\dot\phi=-\frac{dV}{d\phi}
\label{infl3}\ee
where $\phi$ is the inflaton field in a sufficiently homogeneous domain of the
universe.

\subsection{The Hartle-Hawking Wavefunction}

In \cite{ms1,emn2} it was shown, through the energy dependence of quantum
uncertainties, that some sort of stochasticity characterizes non-critical
Liouville string dynamics, implying that the analogy of eq. (\ref{beta2}) with
the equations of motion in inflationary models should be made with those
involving chaotic inflation \cite{lm3}. Let us now briefly review the
properties of these latter models. In such cases, the ground state wavefunction
of the universe may be identified as \cite{hh4}:
\be
\psi_0(a,\phi)=\exp-S_E(a,\phi)
\label{psi04}\ee
where $S_E$ is the Euclidean action for the scalar field $a(\tau)$ and the
inflaton scalar field $\phi(\tau)$ which satisfy the boundary conditions:
\be
a(0)=a~~~~~~,~~~~~~\phi(0)=\phi
\label{bc}\ee
and $\tau$ is the Euclidean time.

To understand how eq. (\ref{psi04}) comes about, we appeal to the
Hartle-Hawking interpretation \cite{hh4}.
Consider the Green's function $\langle
x,t|0,t'\rangle$ of a particle which propagates from the spacetime point
$(0,t')$ to $(x,t)$:
\be
\langle x,t|0,t'\rangle=\sum_n\psi_n^\dagger(x)\psi_n(0)~\e^{iE_n(t-t')}=\int
Dx~\e^{iS(x,|t-t'|)}
\label{green5}\ee
where $\{\psi_n\}$ is the complete set of energy eigenstates with energy
eigenvalues $E_n\geq0$ (the sum in (\ref{green5}) should be replaced by an
appropriate integration in the case of a continuous spectrum).
To obtain an expression for the ground state wavefunction, we make a Wick
rotation $t=-i\tau$, and take the limit $\tau\to-\infty$ to recover the initial
state. Then in the summation over energy eigenvalues in (\ref{green5}), only
the ground state ($n=0$) term survives if $E_0=0$. The corresponding path
integral representation becomes $\int Dx~\e^{-S_E(x)}$, and one obtains eq.
(\ref{psi04}) in the semi-classical approximation.

For inflationary models which are based on the de Sitter spaces $dS_4$ with
\be
a(\tau)=\kappa^{-1}(\phi)\cos\kappa(\phi)\tau
\label{atau}\ee
one has
\be
S_E(a,\phi)=-\frac3{16V(\phi)}
\label{SEinfl}\ee
and hence
\be
\psi_0(a,\phi)=\exp\frac3{16V(\phi)}
\label{psi05a}\ee
Thus the probability density for finding the universe in a state with
$\phi={\rm const.}$, $a=\kappa^{-1}(\phi)=\sqrt{\frac3{8\pi V(\phi)}}$ is
\be
{\cal P}=|\psi_0|^2=\e^{3/8V(\phi)}
\label{P6}\ee
The distribution function (\ref{P6}) has a sharp maximum as $V(\phi)\to0$. For
inflationary models this is a bad feature, because it diminishes the
possibility of finding the universe in a state with a large $\phi$ field and
thereby having a long stage for inflation. However, from the point of view of
Liouville string theory, the result (\ref{P6}), if indeed valid, implies that
the {\it critical} string theory (since $V\propto Q^2$ there) is a favourable
situation statistically, and hence any consideration (such as those in
\cite{ms1}) made in the neighbourhood of a fixed point of the renormalization
group flow on the moduli space of running coupling constants is justified.

\subsection{Moduli Space Wavefunctionals}

Let us now proceed to discuss the possibility of finding a Schr\"odinger wave
equation for the D-particle wavefunction. The identification (\ref{psi04}) in
the inflationary case needs some careful verification in the case of the
topological expansion of the worldsheet $\sigma$-model (\ref{sigma1}). In
Liouville string theory, the genus expansion of the partition function may be
identified \cite{emn2} with the wavefunctional of non-critical string theory in
the moduli space of coupling constants $\{g^I\}$:
\be
\Psi(\{g^I\})=\sum_{\rm genera}\int
Dx~\e^{-S_\sigma[x;\{g^I\}]}\equiv\e^{-{\cal F}[\{g^I\}]}
\label{psig7}\ee
where
\be
{\cal F}[\{g^I\}]=\sum_{h=0}^\infty(g_s)^{h-2}\,{\cal F}_h[\{g^I\}]
\label{effgenusexp}\ee
is the effective target space action functional of
the non-critical string theory. The sum on the right-hand side of
(\ref{effgenusexp}) is over all worldsheet genera, which sums up the
one-particle irreducible connected worldsheet amplitudes ${\cal F}_h$
with $h$ handles. The
gradient flow property (\ref{gflow}) of the $\beta$-functions ensures
\cite{ms1,emn2} that the Helmholtz conditions for canonical quantization are
satisfied, which is consistent with the existence of an off-shell action ${\cal
F}[\{g^I\}]$. In that case, the effective Lagrangian on moduli space whose
equations of motion coincide with the renormalization group equations
(\ref{beta2}) is given by \cite{ms1}
\be
{\cal L}_\mods(t)=-\beta^I\,G_{IJ}\,\beta^J
\label{effcalL}\ee
and it coincides with the Zamolodchikov $C$-function. The semi-classical
wavefunction determined by (\ref{psig7}) is thereby determined by the
action $C[\lambda]$ regarded as an effective action on the space of
two-dimensional renormalizable field theories. Thus the probability density is
${\cal
P}[\{g^I\}]=\e^{-2{\cal F}[\{g^I\}]}$, which implies that the minimization of
${\cal F}[\{g^I\}]$ yields a maximization of ${\cal P}[\{g^I\}]$, provided that
the effective action is positive-definite. This is an ideal situation, since
then the minimization of ${\cal F}[\{g^I\}]$, in the sense of solutions of the
equations $\delta{\cal F}/\delta g^I=0$, corresponds to the
conformally-invariant fixed point of the $\sigma$-model moduli space, thereby
justifying the analysis in a neighbourhood of a fixed point.

However, the identification (\ref{psig7}) is {\it not} the only possibility in
non-critical string theory, as will be discussed below, in particular in
connection with the Schr\"odinger dynamics of D0-branes. The main point is that
upon taking the topological expansion in Liouville string theory, the couplings
$g^I$ become quantized in such a way so that
\be
{\sum_{\rm genera}}'\int
Dx~\e^{-S_\sigma[x;\{g^I\}]}=\int\limits_{\mods(\{g^I\})}D\alpha^I~
\e^{-\frac1{2\Gamma^2}\alpha^IG_{IJ}\alpha^J}\,\int Dx~
\e^{-S_\sigma^{(0)}[x;\{g^I+\alpha^I\}]}
\label{pinchsum8}\ee
where the prime on the sum means that the genus expansion is truncated to a sum
over pinched annuli of infinitesimal strip size, $S_\sigma^{(0)}[x;\{g^I\}]$ is
the tree-level (disc or sphere) action for the $\sigma$-model, and $\alpha^I$
are worldsheet wormhole parameters on the moduli space $\mod(\{g^I\})$ of the
two-dimensional quantum field theory. The Gaussian spread in the $\alpha^I$ in
(\ref{pinchsum8}) can be interpreted as a probability distribution
characterizing the statistical fluctuations of the coupling constants $g^I$.
The width $\Gamma$ is proportional to the logarithmic modular divergences on
the pinched annuli, which may be identified with the short-distance infinities
$\log\Lambda$ at tree-level \cite{ms1} ($\Lambda$ is the worldsheet
ultraviolet cutoff scale). The result (\ref{pinchsum8}) suggests
that one may directly identify the genus expansion of the worldsheet
partition function as the probability density
\be
\left|\Psi(\{g^I\},t)\right|^2\equiv{\cal P}(\{g^I\},t)
\label{PsicalP}\ee
for finding non-critical strings in the moduli space configuration $\{g^I\}$ at
Liouville time $t$ (the worldsheet zero mode of the Liouville field). In this
way one has a {\it natural} explanation for the reality of eq.
(\ref{pinchsum8}) on Euclidean worldsheets. If the identification of the genera
summed partition function with the probability density holds, i.e. with the
square of the wavefunction $\Psi(\{g^I\},t)$ rather than the wavefunctional
itself, then one may obtain a temporal evolution equation for (\ref{PsicalP})
using the Wilson-Polchinski renormalization group equation on the string
worldsheet \cite{hlp6}. This will be described in section IV.

One may argue formally in favour of the above identification 
in the case of Liouville strings, within a world-sheet formalism, 
by noting~\cite{emn} that the 
conventional interpretation of the Liouville (world-sheet) correlators  
as target-space $S$-matrix elements breaks down upon the 
interpretation of the Liouville zero-mode as target time. Instead,
the only well-defined concept in such a case is the non-factorizable 
$\nd{S}$-matrix, which acts on target-space density matrices rather than 
state vectors. This in turn implies that the corresponding 
world-sheet partition function, summed over topologies, which in the case 
of critical strings would be the generating functional of such $S$-matrix
elements in target space, should be identified with the 
probability density in the moduli space of the non-critical strings
(\ref{PsicalP}). 
In the Appendix 
we review this approach~\cite{emn} by focusing on 
those aspects of the formalism that are most relevant to our 
pruposes here. As we shall discuss there, the above identification 
follows from specific properties of the Liouville string formalism.

Notice that if one interprets the topological expansion of the worldsheet
partition function as the probability density for the non-critical string
configuration $\{g^I\}$, then the simple argument leading to eq. (\ref{psi04})
is not valid here. In such a situation the action in eq. (\ref{green5}), which
refers to the string moduli space, is {\it not} the same as the effective
target space action ${\cal F}[\{g^I\}]$, but rather something different,
corresponding to the phase of the wavefunctional $\Psi(\{g^I\},t)$ whose
probability density (\ref{PsicalP}) corresponds to the genera summed worldsheet
partition function. This is not necessarily a bad feature, as we shall see,
although in most treatments the target space effective action ${\cal
F}[\{g^I\}]$ is identified with the moduli space action upon identification of
the Liouville zero mode (i.e. the local worldsheet renormalization group scale)
with target time. For this, we observe that the statistical interpretation of
the resummed worldsheet partition function is {\it compatible} with the
interpretation in \cite{ms1} of the Gaussian wormhole parameter distribution
function in eq. (\ref{pinchsum8}) as being responsible for the quantum
uncertainties of D-branes. This follows trivially from the fact that
\be
\left|\Psi(\{g^I\},t)\right|^2=\e^{-2{\cal F}(\{g^I\},t)}
\label{PsicalF}\ee
Then, any correlation function may be written as
\bea
\left\langle V_{I_1}\cdots
V_{I_n}\right\rangle&=&\int\limits_{\mods(\{g^I\})}Dg^I~
\left|\Psi(\{g^I\},t)\right|^2\,V_{I_1}\cdots V_{I_n}\nonumber\\&=&
\int\limits_{\mods(\{g^I\})}Dg^I~\int\limits_{\mods(\{g^I\})}D\alpha^I~
\e^{-\frac1{2\Gamma^2}\alpha^IG_{IJ}\alpha^J}\,\int Dx~\e^{-S_
\sigma^{(0)}[x;\{g^I+\alpha^I\}]}\,V_{I_1}\cdots V_{I_n}
\label{VIcorr}\eea
which using eq. (\ref{PsicalF}) gives the connection between the two
probability distributions.

\section{Matrix D-brane Dynamics}

In this section we shall briefly review the worldsheet description of
\cite{ms1} for matrix D0-brane dynamics. The partition function is given by
\cite{dorn}
\bea
{\cal Z}[A_0,Y]&=&\int
D\mu(x,\bar\xi,\xi)~\exp\left(-\frac1{4\pi\alpha'}\int\limits_\Sigma
d^2z~\eta_{\mu\nu}\,\partial x^\mu\,\bar\partial x^\nu+\frac1{2\pi\alpha'}
\oint\limits_{\partial\Sigma}d\tau~x_i(\tau)\,\partial_\sigma x^i(\tau)
\right)\nonumber\\& &\times\,{\cal
W}[x,\bar\xi,\xi]
\label{partfnD}\eea
where
\be
{\cal W}[x,\bar\xi,\xi]=\exp
ig_s\oint\limits_{\partial\Sigma}d\tau~\left(\bar\xi_a(\tau)
A_0^{ab}\xi_b(\tau)\,\partial_\tau x^0(\tau)+\frac i{2\pi\alpha'}
\,\bar\xi_a(\tau)Y_i^{ab}(x^0)\xi_b(\tau)\,\partial_\sigma x^i(\tau)\right)
\label{wilsonloop}\ee
is the deformation action of the free $\sigma$-model in (\ref{partfnD}). Here
the indices $\mu=0,1,\dots,9$ and $i=1,\dots,9$ label spacetime and spatial
directions of the target space, which we assume has a flat metric
$\eta_{\mu\nu}$. The functional
integration measure in (\ref{partfnD}) is given by
\be
D\mu(x,\bar\xi,\xi)=Dx^\mu~D\bar\xi~D\xi~\exp\left[-\sum_{a=1}^N
\left(\oint\limits_{\partial\Sigma}d\tau~\bar\xi_a(\tau)\,\partial_\tau
\xi_a(\tau)+\bar\xi_a(0)\xi_a(0)\right)\right]
\,\sum_{a=1}^N\bar\xi_a(0)\xi_a(1)
\label{Dmeasure}\ee
The complex auxilliary fields $\bar\xi_a(\tau)$ and $\xi_a(\tau)$,
$a=1,\dots,N$,
transform in the fundamental representation of the brane gauge group, and they
live on the boundary of the worldsheet $\Sigma$ which at tree-level is a disc
whose boundary is a circle $\partial\Sigma$ with periodic longitudinal
coordinate
$\tau\in[0,1]$ and normal coordinate $\sigma\in\real$. They have the propagator
$\langle\bar\xi_a(\tau)\xi_b(\tau')\rangle=\delta_{ab}\,\Theta(\tau'-\tau)$,
where $\Theta$ denotes the usual step function. The integration over the
auxilliary fields with the measure (\ref{Dmeasure}) therefore turns
(\ref{wilsonloop}) into a path-ordered exponential functional of the fields $x$
which is the $T$-dual of the usual Wilson loop operator for the ten-dimensional
gauge field $(A^0,-\frac1{2\pi\alpha'}Y^i)$ dimensionally reduced to the
D-particle worldlines. In this picture, $A^0$ is thought of as a gauge field
living on the brane worldline, while $Y_i^{aa}$,
$a=1,\dots,N$, are the transverse coordinates of
the $N$ D-particles and $Y_i^{ab}$, $a\neq b$, of the short open string
excitations connecting them. We shall subtract out the center of mass motion of
the
assembly of $N$ D-branes and assume that $Y_i\in su(N)$. We shall also use
$SU(N)$-invariance of the theory (\ref{partfnD}) to select the temporal gauge
$A^0=0$.

The action in (\ref{partfnD}) may be formally identified with the
deformed conformal field theory (\ref{sigma1}) by taking the couplings $g^I\sim
Y_i^{ab}$ and introducing the one-parameter family of bare matrix-valued vertex
operators
\be
V_{ab}^i(x;\tau)=\frac{g_s}{2\pi\alpha'}\,\partial_\sigma
x^i(\tau)\,\bar\xi_a(\tau)\xi_b(\tau)
\label{Dvertexops}\ee
This means that there is a one-parameter family of Dirichlet boundary
conditions for the fundamental string fields $x^i$ on $\partial\Sigma$,
labelled by $\tau\in[0,1]$ and the configuration fields
\be
y_i(x^0;\tau)=\bar\xi_a(\tau)\,Y_i^{ab}(x^0(\tau))\,\xi_b(\tau)
\label{bdryfields}\ee
Instead of being forced to sit on a unique hypersurface as in the case of a
single D-brane, in the non-abelian case there is an infinite set of
hypersurfaces on which the string endpoints are situated. In this sense the
coordinates (\ref{bdryfields}) may be thought of as an ``abelianization'' of
the non-abelian D-particle coordinate fields $Y_i^{ab}$.

To describe the non-relativistic dynamics of heavy D-particles, the natural
choice is to take the couplings to correspond to the Galilean boosted
configurations $Y_i^{ab}(x^0)=Y_i^{ab}+U_i^{ab}x^0$, where $U_i$ is the
non-relativisitic velocity matrix. However, logarithmic modular divergences
appear in matter field amplitudes at higher genera
when the string propagator $L_0$ is computed with
Dirichlet boundary conditions. These modular divergences are cancelled by
adding logarithmic recoil operators \cite{ms1,kmw12} to the matrix
$\sigma$-model action in (\ref{partfnD}). From a physical point of view, if one
is to use low-energy probes to observe short-distance spacetime structure, such
as a generalized Heisenberg microscope, then one needs to consider the
scattering of string matter off the assembly of D-particles. For the
Galilean-boosted multiple
D-particle system, the recoil is described by taking the
deformation of the $\sigma$-model action in (\ref{partfnD}) to be of the form
\cite{ms1}
\be
Y_i^{ab}(x^0)=\sqrt{\alpha'}\,Y_i^{ab}C_\epsilon(x^0)+U_i^{ab}D_\epsilon(x^0)=
\left(\sqrt{\alpha'}\,\epsilon Y_i^{ab}+U_i^{ab}x^0\right)\Theta_\epsilon(x^0)
\label{recoilY}\ee
where
\be
C_\epsilon(x^0)=\epsilon\,\Theta_\epsilon(x^0)~~~~~~,~~~~~~
D_\epsilon(x^0)=x^0\,\Theta_\epsilon(x^0)
\label{recoilops}\ee
and
\be
\Theta_\epsilon(x^0)=\frac1{2\pi
i}\int\limits_{-\infty}^{+\infty}\frac{dq}{q-i\epsilon}~\e^{iqx^0}
\label{stepfnreg}\ee
is the regulated step function whose $\epsilon\to0^+$ limit is the usual step
function. The operators (\ref{recoilops}) have non-vanishing matrix elements
between different string states and therefore describe the appropriate change
of quantum
state of the D-brane background. They can be thought of as describing the
recoil of the assembly of D-particles in an impulse approximation, in which it
starts moving as a whole only at time $x^0=0$. The collection of constant
matrices $\{Y^i_{ab},U^j_{cd}\}$ now form the set of coupling constants
$\{g^I\}$ for the
worldsheet $\sigma$-model (\ref{partfnD}).

The recoil operators (\ref{recoilops}) possess a very important property. They
lead to a deformation of the free $\sigma$-model action in (\ref{partfnD})
which is not conformally-invariant, but rather defines a logarithmic conformal
field theory \cite{gurarie}.
Such a quantum field theory contains logarithmic scaling violations in its
correlation functions on the worldsheet, which can be seen in the present case
by computing the pair correlators of the fields (\ref{recoilops})
\cite{kmw12}
\bea
\Bigl\langle
C_\epsilon(z)\,C_\epsilon(0)\Bigr\rangle&=&0\nn\\\Bigl\langle
C_\epsilon(z)\,D_\epsilon(0)\Bigr\rangle&=&\frac
b{z^{h_\epsilon}}\nn\\\Bigl\langle
D_\epsilon(z)\,D_\epsilon(0)\Bigr\rangle&=&\frac{b\,\alpha'}
{z^{h_\epsilon}}\,\log z
\label{2ptconfalg}\eea
where
\be
h_\epsilon=-\frac{|\epsilon|^2\,\alpha'}2
\label{CDconfdim}\ee
is the conformal dimension of the recoil operators. The constant $b$ is fixed
by the leading logarithmic divergence of the conformal blocks of the theory.
Note that (\ref{CDconfdim}) vanishes as $\epsilon\to0$, so that the logarithmic
worldsheet divergences in (\ref{2ptconfalg}) cancel the modular annulus
divergences mentioned above. An essential ingredient for this cancellation is
the identification \cite{kmw12}
\be
\frac1{\epsilon^2}=-2\alpha'\log\Lambda
\label{epLambdaid}\ee
which relates the target space regularization parameter $\epsilon$ to the
worldsheet ultraviolet cutoff scale $\Lambda$.

Logarithmic conformal field theories are characterized by the fact that their
Virasoro generator $L_0$ is not diagonalizable, but rather admits a Jordan cell
structure. Here the operators (\ref{recoilops}) form the basis of a $2\times2$
Jordan block and they appear in the spectrum of the two-dimensional quantum
field theory as a consequence of the zero modes that arise from the breaking of
the target space translation symmetry by the topological defects. The mixing
between $C$ and $D$ under a conformal transformation of the worldsheet can be
seen explicitly by considering a scale transformation
\be
\Lambda\to\Lambda'=\Lambda\,\e^{-t/\sqrt{\alpha'}}
\label{Lambdatransf}\ee
Using (\ref{epLambdaid}) it follows that the operators (\ref{recoilops}) are
changed according to $D_\epsilon'=D_\epsilon+t\sqrt{\alpha'}C_\epsilon$ and
$C_\epsilon'=C_\epsilon$. Thus in order to maintain scale-invariance of the
theory (\ref{partfnD}) the coupling constants must transform under
(\ref{Lambdatransf}) as \cite{kmw12,lm11} $Y'^i=Y^i+U^it$ and $U'^i=U^i$,
which are just the Galilean transformation laws for the positions $Y^i$ and
velocities $U^i$. Thus a scale transformation of the worldsheet is equivalent
to a
Galilean transformation of the moduli space of $\sigma$-model couplings, with
the parameter $\epsilon^{-2}$ identified with the time evolution parameter
$t=-\sqrt{\alpha'}\log\Lambda$. The corresponding
$\beta$-functions for the worldsheet renormalization group flow are
\bea
\beta_{Y_i}&\equiv&\frac{dY_i}{dt}=h_\epsilon\,Y_i+\sqrt{\alpha'}\,U_i
\nn\\\beta_{U_i}&\equiv&\frac{dU_i}{dt}=h_\epsilon\,U_i
\label{betafns}\eea
and they generate the Galilean group $G(9)^{N^2}$ in nine-dimensions.

The associated Zamolodchikov metric
\be
G_{ab;cd}^{ij}=2N\Lambda^2\left\langle
V_{ab}^i(x;0)\,V_{cd}^j(x;0)\right\rangle
\label{ZamD}\ee
can be evaluated to leading orders in $\sigma$-model perturbation theory using
the logarithmic conformal algebra (\ref{2ptconfalg}) and the propagator of the
auxilliary fields to give \cite{ms1}
\bea
G_{ab;cd}^{ij}&=&\frac{4\bar g_s^2}{\alpha'}\left[\eta^{ij}\,I_N\otimes
I_N+\frac{\bar g_s^2}{36}\left\{I_N\otimes\left(\bar U^i\bar U^j+\bar U^j\bar
U^i\right)\right.\right.\nn\\& &\biggl.\left.+\,\bar U^i\otimes\bar U^j+\bar
U^j\otimes\bar U^i+\left(\bar U^i\bar U^j+\bar U^j\bar U^i\right)\otimes
I_N\right\}\biggr]_{db;ca}+{\cal O}\left(\bar g_s^6\right)
\label{zammetricexpl}\eea
where $I_N$ is the identity operator of $SU(N)$ and we have introduced the
renormalized coupling constants
\be
\bar g_s=g_s/\sqrt{\alpha'}\epsilon~~~~~~,~~~~~~\bar
U^i=U^i/\sqrt{\alpha'}\epsilon
\label{rencc}\ee
{}From the renormalization group equations (\ref{betafns}) it follows that the
renormalized velocity operator in target space is truly marginal,
\be
\frac{d\bar U^i}{dt}=0
\label{dUdt}\ee
which ensures uniform motion of the D-branes. It can also be shown that the
renormalized string coupling $\bar g_s$ is time-independent \cite{ms1}. If we
further define the position renormalization
\be
\bar Y^i=Y^i/\sqrt{\alpha'}\epsilon
\label{renY}\ee
then the $\beta$-function equations (\ref{betafns}) coincide with the Galilean
equations of motion of the D-particles, i.e.
\be
\frac{d\bar Y^i}{dt}=\bar U^i
\label{dYdt}\ee
Note that the Zamolodchikov metric (\ref{zammetricexpl}) is a complicated
function of the D-brane dynamical parameters, and as such it represents the
appropriate effective target space geometry of the D-particles. The moduli
space Lagrangian (\ref{effcalL}) is then readily seen to coincide with the
expansion to ${\cal O}(\bar g_s^4)$ of the symmetrized form of the non-abelian
Born-Infeld action for the D-brane dynamics \cite{tseytlin},
\be
{\cal L}_{\rm NBI}=\frac1{\sqrt{2\pi\alpha'}\bar g_s}\,\tr~{\rm
Sym}\,\sqrt{\det_{\mu,\nu}\left[\eta_{\mu\nu}\,I_N+2\pi\alpha'\bar
g_s^2\,F_{\mu\nu}\right]}
\label{NBIaction}\ee
where tr denotes the trace in the fundamental representation of $SU(N)$,
\be
{\rm Sym}(M_1,\dots,M_n)=\frac1{n!}\sum_{\pi\in S_n}M_{\pi_1}\cdots
M_{\pi_n}
\label{symprod}\ee
is the symmetrized matrix product and the components of the
dimensionally reduced field strength tensor are given by
\be
F_{0i}=\frac1{2\pi\alpha'}\,\frac{d{\bar Y}_i}{dt}~~~~~~,~~~~~~
F_{ij}=\frac{\bar g_s}{(2\pi\alpha')^2}\left[\bar Y_i\,,\,\bar Y_j\right]
\label{Fred}\ee

\section{Evolution Equation for the Probability Distribution}

In this section we will derive the temporal evolution equation for the
probability density ${\cal P}(\{g^I\},t)$ following the identification of time
with a worldsheet renormalization group scale (i.e. the Liouville zero mode).
The basic identity is the Wilson-Polchinski equation for the case of the
worldsheet action (\ref{sigma1}) which reads \cite{hlp6}
\bea
0&=&\frac{\partial{\cal Z}}{\partial\log\Lambda}\nonumber\\&=&\int
Dx^\mu~\e^{-S_\sigma[x;\{g^I\}]}\,\left\{\frac{\partial S_{\rm
int}}{\partial\log\Lambda}\right.\nonumber\\& &\left.-\,\int\limits_\Sigma
d^2z~\int\limits_\Sigma
d^2w~\left(\frac\partial{\partial\log\Lambda}G(z-w)\right)\left[
\frac{\delta^2S_{\rm int}}{\delta x^\mu(z)\delta x_\mu(w)}+
\frac{\delta S_{\rm int}}{\delta x^\mu(z)}\frac{\delta S_{\rm int}}
{\delta x_\mu(w)}\right]\right\}
\label{WPeq9}\eea
and it is the requirement of conformal invariance of the quantum string theory.
Here $S_{\rm int}=S_\sigma-S_*$, $\cal Z$ is the partition function of the
$\sigma$-model, and
\be
G(z-w)=\left\langle\NO x^\mu(z)x_\mu(w)\NO\right\rangle_*
\label{2ptfn}\ee
is the two-point function computed with respect to the conformal field theory
action $S_*[x]$. The basic assumption in arriving at eq. (\ref{WPeq9}) is that
the ultra-violet cutoff $\Lambda$ on the string worldsheet appears explicitly
only in the propagator $G(z-w)$, as can always be arranged by an appropriate
regularization \cite{hlp6}.

Henceforth we shall concentrate on the specific case of interest of a system of
$N$ interacting D-particles. Then, upon summing up over pinched genera, there
are extra logarithmic divergences in the Green's function (\ref{2ptfn}) coming
from pinched annulus diagrams, which may be removed by the introduction of
logarithmic recoil operators, as explained in the previous section. Using
primes to denote the result of resumming the topological expansion over pinched
genera, we then have that
\be
\frac{\partial}{\partial\log\Lambda}G(z-w)'=
\frac\partial{\partial\log\Lambda}{\sum_{\rm genera}}'\left\langle\NO
x^\mu(z)x_\mu(w)\NO\right\rangle=\frac\partial{\partial\log\Lambda}
\left\langle\NO x^\mu(z)x_\mu(w)\NO\right\rangle_{\rm int}
\label{corrint}\ee
where the correlator $\langle\cdot\rangle_{\rm int}$ includes the disc and
recoil interaction contributions. Subtracting the disc $\Lambda$-dependence in
normal ordering, the remaining dependence on the worldsheet cutoff comes from
the two-point functions of the logarithmic recoil operators, giving terms of
the form
\be
\frac\partial{\partial\log\Lambda}\left\langle\NO
x^\mu(z)x_\mu(w)\NO\Bigl(a_{CC}C_\epsilon(z)C_\epsilon(w)+a_{CD}
C_\epsilon(z)D_\epsilon(w)+a_{DD}D_\epsilon(z)D_\epsilon(w)\Bigr)
\right\rangle_*
\label{recoilcontrtot}\ee
The leading divergence comes from the correlation function $\langle
D_\epsilon(z)D_\epsilon(w)\rangle_*\sim\log\Lambda$, which follows upon the
identification (\ref{epLambdaid}). Thus we may write
\be
\frac\partial{\partial\log\Lambda}G(z-w)\simeq
c_G\,(\alpha')^2\log|z-w|\,\sum_{i=1}^9\,\sum_{a,b=1}^N|U_{ab}^i|^2
\label{leadingGLambda}\ee
where $c_G>0$ is a numerical coefficient whose precise value is not important,
and we have used the fact that $U^i\in su(N)$.

Next, we observe that in the case of D-particles the second term in eq.
(\ref{WPeq9}) becomes
\bea
& &\int
D\mu(x,\bar\xi,\xi)~\e^{-S_\sigma}\,\oint\limits_{\partial\Sigma}d\tau~\oint
\limits_{\partial\Sigma}d\tau'~(\alpha')^2c_G\sum_{i=1}^9\sum_{a,b=1}^N
|U_{ab}^i|^2\,\log[2-2\cos(\tau-\tau')]\nonumber\\& &~~~~~~\times\,
\left[\frac{\delta^2S_{\rm int}}{\delta x^\mu(\tau)\delta x_\mu(\tau')}
+\frac{\delta S_{\rm int}}{\delta x^\mu(\tau)}
\frac{\delta S_{\rm int}}{\delta x_\mu(\tau')}\right]
\label{2ndterm}\eea
where the interaction Lagrangian is given by
\be
S_{\rm
int}=\frac{g_s}{2\pi\alpha'}\oint\limits_{\partial\Sigma}d\tau~\partial_\sigma
x^i(\tau)\,\bar\xi_a(\tau)\,Y_i^{ab}(x^0)\,\xi_b(\tau)
\label{DSint}\ee
In the case of a system of recoiling D0-branes, the $\sigma$-model couplings in
eq. (\ref{DSint}) are given by (\ref{recoilY}) with the abelianized couplings
(\ref{bdryfields}) of $Y_i^{ab}$ viewed as the boundary values for the open
string embedding fields $x^i(\tau)$ on the D-brane. This means that the fields
$x^i(\tau)$ are simply identified with $\bar\xi_a(\tau)Y_i^{ab}\xi_b(\tau)$.
All the non-trivial dependence comes from the $x^0$ field which obeys Neumann
boundary conditions and is not constant on the boundary of $\Sigma$. Then we
may write
\bea
& &\frac{\delta^2S_{\rm int}}{\delta x^\mu(\tau)\delta
x_\mu(\tau')}+\frac{\delta S_{\rm int}}{\delta x^\mu(\tau)}\frac{\delta S_{\rm
int}}{\delta x_\mu(\tau')}\nonumber\\& &~~~~=\nabla_{y_i}^2S_{\rm
int}+\left(\nabla_{y_i}S_{\rm
int}\right)^2+\left(\frac{g_s}{2\pi\alpha'}\right)^2\,U_i^{ab}U_j^{cd}
\bar\xi_a(\tau)\xi_b(\tau)\bar\xi_c(\tau')\xi_d(\tau')\nonumber
\\& &~~~~~~\times\,\partial_\sigma x^i(\tau)\partial_\sigma x^j(\tau')
\Theta_\epsilon(x^0(\tau))\Theta_\epsilon(x^0(\tau'))
\label{Sintderivs}\eea
where $y_i$ denotes the constant abelianized zero modes of $x^i(\tau)$ on
$\partial\Sigma$. Here we have used the fact that terms of the form
$x^0\delta(x^0)$ and $\Theta(x^0)\delta(x^0)$ vanish with the regularization
(\ref{stepfnreg}) \cite{ms1}. The terms involving $\partial_\sigma
x^i(\tau)\partial_\sigma x^j(\tau')$ will average out to yield terms of the
form
\be
|U_{ab}^i|^2\left\langle\Theta_\epsilon(x^0(\tau))\Theta_\epsilon(x^0(\tau'))
\right\rangle_*=\alpha'|\bar U_{ab}^i|^2\left\langle C_\epsilon(\tau)
C_\epsilon(\tau')\right\rangle_*\sim{\cal O}(\epsilon^2)
\label{Oepterms9b}\ee
where we have used the logarithmic conformal algebra. At leading orders, these
terms vanish, but we shall see the importance of such sub-leading terms later
on.

Using the Dirichlet correlator
\be
\left\langle\partial_\sigma x^i(\tau)\partial_\sigma
x_i(\tau')\right\rangle_*=-\frac{36\pi^2\alpha'}{1-\cos(\tau-\tau')}
\label{Dcorr}\ee
we find that the boundary integrations in eq. (\ref{2ndterm}) are of the form
\cite{ms1}
\be
\oint\limits_{\partial\Sigma}d\tau~\oint\limits_{\partial\Sigma}d\tau'~
\frac{\log[2-2\cos(\tau-\tau')]}{1-\cos(\tau-\tau')}\sim\log\Lambda
\label{bdryints}\ee
which has the effect of renormalizing the velocity matrix $U_{ab}^i\to\bar
U_{ab}^i$. Thus, ignoring the ${\cal O}(\epsilon^2)$ terms for the moment, we
find that the remaining terms in the Wilson-Polchinski renormalization group
equation (\ref{WPeq9}) yield a diffusion term for the probability density:
\be
\partial_t{\cal P}[Y,U;t]=c_G\sqrt{\alpha'}\sum_{j=1}^9\sum_{a,b=1}^N|\bar
U_{ab}^j|^2\,\nabla_{y_i}^2{\cal P}[Y,U;t]+{\cal O}(\epsilon^2)
\label{diffprob10}\ee
This equation is of the Fokker-Planck type,
with diffusion coefficient
\be
{\cal D}=c_G\sqrt{\alpha'}\sum_{i=1}^9\sum_{a,b=1}^N|\bar U_{ab}^i|^2
\label{diffcoeff11}\ee
coming from the quantum recoil of the assembly of D-particles. The diffusion
disappears when there is no recoil. Note that (\ref{diffcoeff11}) naturally
incorporates the short-distance quantum gravitational smearings for the open
string interactions (compare with eq. (\ref{Deltayia})), and it arises as an
abelianized velocity for the constant auxilliary field configuration
$\bar\xi_a(\tau)=\xi_a(\tau)=1~~\forall a=1,\dots,N$.

The evolution equation (\ref{diffprob10}) should be thought of as a
modification of the usual continuity equation for the probability density.
Indeed, as we will now show, the ${\cal O}(\epsilon^2)$ terms in eq.
(\ref{diffprob10}) coming from (\ref{Oepterms9b}) are of the form
$-\nabla_{y_i}{\cal J}_i$, where
\be
{\cal
J}_i=\frac{\hbar_\mods}{2im}\left(\Psi^\dagger\,\nabla_{y_i}\Psi-
\Psi\,\nabla_{y_i}\Psi^\dagger\right)
\label{current13}\ee
is the probability current density. Here
\be
m=\frac1{\sqrt{\alpha'}\bar g_s}~~~~~~,~~~~~~\hbar_\mods=4\bar g_s
\label{mhbardefs}\ee
are, respectively, the
BPS mass of the D-particles and the moduli space
``Planck constant''.~\footnote{\baselineskip=12pt The
identification (\ref{mhbardefs}) of
Planck's constant in the D-particle quantum mechanics on moduli space
with the string coupling constant is actually not unique
in the present context of considering only the exchange of strings
between D-particles. As discussed in \cite{ms1}, the most general
relation, compatible with the
logarithmic conformal algebra, involves an arbitrary exponent $\chi$ through
$\hbar_\mods=4(\bar g_s)^{1 + \chi/2}$. The exponent $\chi$
arises from specific mechanisms for the cancellation of modular
divergences on pinched annular surfaces by appropriate
world-sheet short-distance
infinities at lower genera. The only restriction imposed
on $\chi$ is that it be positive definite.
As shown in \cite{ms1}, the standard kinematical properties of D-particles
are reproduced by the choice $\chi=\frac23$.
A choice of $\chi\ne0$ seems more natural from the point of view
that modular divergences should be suppressed for
weakly interacting strings. However, in the present case,
we assume for simplicity the value $\chi=0$, which yields the standard
string smearing $\sqrt{\alpha '}$ for the minimum length uncertainty.
The incorporation of an arbitrary $\chi\ge 0$ in the formalism
is straightforward and would not affect the qualitative properties of
the following results.}

For this, we note first of all that such terms should generically come in the
form
\be
-\nabla_{y_i}{\cal
J}_i=-\frac{\hbar_\mods}m\left(\nabla_{y_i}^2\,{\rm arg}\,\Psi\right){\cal
P}-\frac{\hbar_\mods}m\left(\nabla_{y_i}\,{\rm
arg}\,\Psi\right)\nabla_{y_i}{\cal P}
\label{genterms}\ee
The second term in eq. (\ref{genterms}), upon
identification of the probability density $\cal P$ with the genera resummed
partition function on the string worldsheet, is proportional to the worldsheet
renormalization group $\beta$-function, given the gradient flow property
(\ref{gflow}) of the string effective action \cite{mm3b}, so that
\be
\nabla_{y_i}{\cal P}=-2{\cal P}\,G_{ij}\beta^j
\label{gflowcalP}\ee
which is to be understood in terms of abelianized quantities. In the present
case the renormalization group equations are given by (\ref{dUdt}) and
(\ref{dYdt}) and, since the couplings $\bar U_i^{ab}$ are truly marginal, we
are left in (\ref{gflowcalP}) with only a Zamolodchikov metric contribution
$G_{CC}=2N\Lambda^4\langle C_\epsilon(\tau)C_\epsilon(\tau)\rangle_*$ (Note
that here one should use suitably normalized correlators
$\langle\cdot\rangle_*$
which yield the behaviour (\ref{gflowcalP})). From the logarithmic conformal
algebra it therefore follows that a term with the structure of the second piece
in eq. (\ref{genterms}) is hidden in the contributions (\ref{Oepterms9b}) which
were dropped as being subleading in $\epsilon$. Furthermore, from
(\ref{zammetricexpl}), (\ref{genterms}), (\ref{leadingGLambda}),
(\ref{Sintderivs}) and (\ref{Oepterms9b}) it follows that to leading orders
\be
\nabla_{y_i}\,{\rm arg}\,\Psi=-\frac{c_G}{\sqrt{\alpha'}\bar
u_i}\left(\sum_{j=1}^9\sum_{a,b=1}^N|\bar U_{ab}^j|^2\right)^2
\label{thetaeq}\ee
where $\bar u_i=d\bar y_i/dt$ is the worldsheet zero mode of the abelianized,
renormalized velocity operator. It then follows that to leading orders we have
$\nabla_{y_i}^2\,{\rm arg}\,\Psi=0$.\footnote{\baselineskip=12pt
Noncommutative position
dependent terms arising from commutators $[Y_i,Y_j]$ appear
only at two-loop order in $\sigma$-model perturbation theory \cite{ms1}. An
interesting extension of the present analysis would be to generalize the
results to include these higher-order terms into the quantum dynamics. However,
given that the pertinent equations involve only the abelianized coordinates
(\ref{bdryfields}), we do not expect the inclusion of such terms to affect the
ensuing qualitative conclusions. The effect of the noncommutativity is to
render the quantum wave equation for the system of D-particles non-linear,
through the recoil-induced diffusion from the multi-brane interactions, as we
discuss in the subsequent sections (for a single brane one would obtain a free
wave equation governing the quantum dynamics).}

Thus, keeping the subleading terms in the target space regularization parameter
$\epsilon$ leads to the complete Fokker-Planck equation for the probability
density ${\cal P}=\Psi^\dagger\Psi$:
\be
\partial_t{\cal P}[Y,U;t]=-\nabla_{y_i}{\cal J}_i[Y,U;t]+{\cal
D}\,\nabla_{y_i}^2{\cal P}[Y,U;t]
\label{FPeq12}\ee
where ${\cal J}_i$ is the probability current density (\ref{current13}), with
$\Psi$ the wavefunctional for the system of D-branes:
\be
\Psi[Y,U;t]=\prod_{i=1}^9\exp\left[-\frac{ic_G}{\sqrt{\alpha'}}\frac{y_i}{\bar
u_i}\left(\sum_{j=1}^9\tr\,|\bar
U^j|^2\right)^2\right]\,\Bigl|\Psi[Y,U;t]\Bigr|
\label{Psi14}\ee
Such quantum diffusion is characteristic of all Liouville string theories
\cite{emn2,emn10,emn10b}. The resulting quantum dynamics, including the
quantum diffusion which arises from the D-brane recoil, is described by the
Schr\"odinger wave equation which corresponds to this Fokker-Planck equation.
This equation is analysed in detail in the next section.

\section{Non-linear Schr\"odinger Wave Equations}

Given the Fokker-Planck equation (\ref{FPeq12}), there is no unique solution
 for the wavefunction $\Psi$, as we discuss below, and the resulting
Schr\"odinger wave equation is necessarily non-linear, due to the diffusion
term \cite{dg7,dg8}. Consider the quantum mechanical system with diffusion
which is described by the
Fokker-Planck equation (\ref{FPeq12}) for the probability density ${\cal
P}=\Psi^\dagger\Psi$. In \cite{dg7} it was shown that, by imposing
diffeomorphism invariance in the space $\vec y\in\mod$ and representing the
symmetry through the infinite-dimensional kinematical symmetry algebra
$C^\infty(\mod)~\semiplus{\rm Vect}(\mod)$, one may arrive
at the following {\it non-linear} Schr\"odinger wave equation:
\be
i\hbar_\mods\,\frac{\partial\Psi}{\partial t}={\cal H}_0\Psi+iI(\Psi)\Psi
\label{nlse16a}\ee
where ${\cal H}_0$ is the linear Hamiltonian operator
\be
{\cal H}_0=-\frac{\hbar_\mods^2}{2m}\,\nabla_{y_i}^2+V_\mods(\vec y,\vec u;t)
\label{Ham16b}\ee
and
\be
I(\Psi)=\frac12\,\hbar_\mods\,{\cal
D}\,\frac{\nabla_{y_i}^2(\Psi^\dagger\Psi)}{\Psi^\dagger\Psi}
\label{IPsi}\ee
Here $V_\mods(\vec y,\vec u;t)$ is the interaction potential on moduli space
and the real continuous quantum number $\cal D$ in (\ref{diffcoeff11}) is the
classification parameter of the unitarily inequivalent diffeomorphism group
representations. Other models which have more than one type of diffusion
coefficient
can be found in \cite{dg7,dg8}.

A crucial point \cite{dg8} is that there exist non-linear phase
transformations of the wavefunction $\Psi$ (known as quantum mechanical ``gauge
transformations'') which leave invariant appropriate families of
non-linear Schr\"odinger equations, and also the probability density $\cal P$.
Such transformations do not affect any physical observables of the system. This
implies that the choice of $\Psi$ is ambiguous, once a density $\cal P$ is
found as a solution of eq. (\ref{FPeq12}) on the collective coordinate space
$\{Y_i^{ab}\}$ of the D-branes. An important ingredient in finding such
transformations is the assumption \cite{dg8,fh9} that all measurements of
quantum mechanical systems can be made so as to reduce eventually to position
and time measurements. Because of this possibility, a theory formulated in
terms of position measurements is complete enough in principle to describe all
quantum phenomena. This point of view is certainly met by the D-brane moduli
space, whereby the wavefunctional depends only on the couplings $\{g^I\}$ and
not on the conjugate momenta $p_I=-i\hbar_\mods\,\partial/\partial g^I$.
The group of non-linear gauge transformations acts on each leaf in a foliation
of a family of non-linear Schr\"odinger equations, such that the
two-dimensional leaves of the foliation consist of sets of equivalent
quantum mechanical evolution equations.

It follows that then one can perform the following local, two-parameter
projective gauge transformation of the wavefunction \cite{dg8}:
\be
\Psi'=N_{\gamma,\lambda}(\Psi)=|\Psi|\exp(i\gamma\log|\Psi|+i\lambda\,{\rm
arg}\,\Psi)
\label{phasetransf17}\ee
under which the probability density is invariant, but the probability
current transforms as
\be
{\cal J}_i'=\lambda\,{\cal J}_i+\frac\gamma2\,\nabla_{y_i}{\cal P}
\label{probcurrenttransf}\ee
Here $\gamma(t)$ and $\lambda(t)\neq0$ are some real-valued time-dependent
functions. The collection of all non-linear transformations
$N_{\gamma,\lambda}$
obeys the multiplication law of the one-dimensional affine Lie group
$Aff(1)$. Under (\ref{phasetransf17})
there are families of non-linear Schr\"odinger equations that are
{\it closed} (in the sense of ``gauge closure''). A generic form of such a
family, to which the non-linear Schr\"odinger equation (\ref{nlse16a}) belongs,
is
\bea
i\,\frac{\partial\Psi}{\partial
t}&=&\frac1{\hbar_\mods}\,{\cal
H}_0\Psi+i\nu_2R_2[\Psi]\,\Psi+\mu_1R_1[\Psi]\,\Psi+\left(\mu_2-
\mbox{$\frac12$}\,\nu_1\right)R_2[\Psi]\,\Psi\nn\\& &+\,(\mu_3+\nu_1)
R_3[\Psi]\,\Psi+\mu_4R_4[\Psi]\,\Psi+\left(\mu_5+\mbox{$\frac14$}\,\nu_1
\right)R_5[\Psi]\,\Psi\nn\\&=&i\sum_{i=1,2}\nu_iR_i[\Psi]\,\Psi+
\sum_{j=1}^5\mu_jR_j[\Psi]\,\Psi +\frac1{\hbar_\mods}\,
V_\mods(\vec y,\vec u;t)\,\Psi
\label{family18}\eea
where $\nu_i,\mu_j$ are real-valued coefficients which are related to diffusion
coefficients $\cal D$ and ${\cal D}'$ by
\bea
\nu_1&=&-\frac{\hbar_\mods}{2m}\nn\\\nu_2&=&\frac12\,{\cal
D}\nn\\\mu_1&=&c_1{\cal D}'\nn\\\mu_2&=&-\frac{\hbar_\mods}{4m}+c_2{\cal
D}'\nn\\\mu_3&=&\frac{\hbar_\mods}{2m}+c_3{\cal D}'\nn\\\mu_4&=&c_4{\cal
D}'\nn\\\mu_5&=&\frac{\hbar_\mods}{8m}+c_5{\cal D}'
\label{numu19}\eea
and $R_j[\Psi]$ are non-linear homogeneous functionals of degree 0 which are
defined by
\bea
R_1&=&\frac m{\hbar_\mods}\,\frac{\nabla_{y_i}{\cal J}_i}{\cal
P}\nn\\R_2&=&\frac{\nabla_{y_i}^2{\cal P}}{\cal
P}\nn\\R_3&=&\frac{m^2}{\hbar_\mods^2}\,\frac{{\cal J}_i^2}{{\cal
P}^2}\nn\\R_4&=&\frac m{\hbar_\mods}\,\frac{{\cal J}_i\,\nabla_{y_i}{\cal
P}}{{\cal P}^2}\nn\\R_5&=&\frac{(\nabla_{y_i}{\cal P})^2}{{\cal P}^2}
\label{Rj20}\eea
In eq. (\ref{numu19}) the $c_j$ are constants, while in eq. (\ref{Rj20}) the
probability current density is given by (\ref{current13}) with ${\cal
P}=\Psi^\dagger\Psi$.

The gauge group $Aff(1)$ acts on the parameter space of the family
(\ref{family18}). Some members of this family are thereby linearizable to an
ordinary Schr\"odinger wave equation under the action of (\ref{phasetransf17}).
These are the members for which there exists a specific relation between $\cal
D$ and ${\cal D}'$ \cite{dg8}, and for which Ehrenfest's theorem of quantum
mechanics receives no dissipative corrections. The quantum mechanics of
D-particles is not of this type, given that there is definite diffusion,
dissipation and thus time irreversability.
However, as discussed in \cite{ms1,kmw12,lm11}, one needs to also maintain
Galilean invariance, which is a property originating from the logarithmic
conformal algebra of the recoil operators. As described in \cite{dg8}, there is
a class of non-linear Schr\"odinger wave equations which is Galilean invariant
but which violates time-reversal symmetry. For this, it is useful to first
construct a parameter set of equations of the form (\ref{numu19}) which remain
{\it invariant} under the gauge transformations (\ref{phasetransf17}). We may
describe the parameter family of equations (\ref{family18}) in terms of orbits
of $Aff(1)$ by regarding $\gamma=2m\mu_1$ and $\lambda=2m\nu_1$ as the group
parameters of an $Aff(1)$ gauge transformation (\ref{phasetransf17}).
Then the remaining five parameters in (\ref{numu19}) are taken to be
the functionally-independent parameters
$\eta_j$,
$j=1,\dots,5$, which are invariant under $Aff(1)$ and are defined by
\bea
\eta_1&=&\nu_2-\frac12\,\mu_1\nn\\\eta_2&=&\nu_1\mu_2-\nu_2\mu_1\nn\\
\eta_3&=&\frac{\mu_3}{\nu_1}\nn\\\eta_4&=&\mu_4-\mu_1\,\frac{\mu_3}{\nu_1}
\nn\\\eta_5&=&\nu_1\mu_5-\nu_2\mu_4+(\nu_2)^2\,\frac{\mu_3}{\nu_1}
\label{etaj21}\eea
A detailed discussion of the corresponding physical observables is given in
\cite{dg8}. For our purposes, we simply select the following relevant property
of the non-linear Schr\"odinger equation based on the parameter set
(\ref{etaj21}).

Consider the effect of time-reversal on the non-linear Schr\"odinger wave
equation. Setting $t\to-t$ is equivalent to introducing the following new set
of coefficients:
\bea
(\nu_i)^T&=&-\nu_i~~~~i=1,2\nn\\(\mu_j)^T&=&-\mu_j~~~~j=1,\dots,5\nn\\
(V_\mods)^T&=&-V_\mods
\label{Tcoeffs22}\eea
where the superscript $T$ denotes the time-reversal transformation. It is
straightforward to show \cite{dg8} that, in terms of the $\eta_j$'s, there is
time-reversal invariance in the non-linear Schr\"odinger equation if
the two parameters $\eta_1$ and $\eta_4$ are both non-vanishing. On the
other hand, a straightforward calculation also shows \cite{dg8} that Galilean
invariance sets $\eta_4=0$, thereby implying that a family of
non-linear Schr\"odinger wave equations which is invariant under $G(9)$ but not
time-reversal invariant indeed {\it exists}. For a single diffusion coefficient
${\cal D}\neq0$, as in the case (\ref{diffcoeff11}) of recoiling D-branes, one
may set ${\cal D}'c_j=0$ (corresponding to the $Aff(1)$ gauge choice $\mu_1=0$)
and thereby obtain the set of gauge invariant
parameters:
\bea
\eta_1&=&\frac12\,{\cal
D}\nn\\\eta_2&=&2\alpha'\bar g_s^4\nn\\\eta_3&=&-1\nn\\\eta_4&=&0\nn\\
\eta_5&=&-\alpha'\bar g_s^4-\frac14\,{\cal D}^2
\label{etainv23}\eea
The parameter set (\ref{etainv23}) breaks time-reversal invariance, as expected
from the non-trivial entropy production and decoherence characterizing the
worldsheet renormalization group approach to target space time involving
Liouville string theory \cite{ms1,emn2,emn13}.
But it {\it does} preserve Galilean
invariance, as is required by conformal invariance of the non-relativistic,
recoiling system of D-particles.

One may therefore propose that the Fokker-Planck equation for the probability
density $\cal P$ on the moduli space of collective coordinates of a system of
interacting D-branes implies a Schr\"odinger wave equation for the pertinent
wavefunctional which is non-linear, Galilean-invariant and has a time arrow,
corresponding to entropy production, and hence explicitly broken time-reversal
invariance. The existence of a dissipation ${\cal D}\propto\tr\,|\bar U^i|^2$,
due to the quantum recoil of the D-branes, implies that the Ehrenfest relations
acquire extra dissipative terms for this family of non-linear Schr\"odinger
equations. For example, one can immediately obtain the relations \cite{dg7}
\bea
\frac
d{dt}\Bigl\langle\!\!\Bigl\langle\widehat{p}_i\Bigr\rangle\!\!\Bigr\rangle
&=&-\Bigl\langle\!\!\Bigl\langle\nabla_{y_i}
V_\mods\Bigr\rangle\!\!\Bigr\rangle-m\int\limits_\mods d\vec y~\Psi^\dagger
\left(\frac{{\cal J}_i^{({\cal D}=0)}}{\cal P}\right)
\left(-\frac{{\cal D}\,\nabla_{y_j}^2{\cal P}}{\cal P}\right)\Psi\nn\\&
 &+\,m\int\limits_\mods d\vec y~\Psi^\dagger\left(-\frac{{\cal D}\,
\nabla_{y_i}{\cal P}}{\cal P}\right)\left(\frac{\nabla_{y_j}
{\cal J}_j^{({\cal D}=0)}}{\cal P}\right)\Psi\nn\\\frac d{dt}
\left\langle\!\!\left\langle\widehat{\bar
y}_i\right\rangle\!\!\right\rangle&=&\left\langle\!\!\left\langle\widehat{\bar
u}_i\right\rangle\!\!\right\rangle
\label{ehren24}\eea
where ${\cal J}_i^{({\cal D}=0)}$ is the undissipative current density
(\ref{current13}). Note that the fundamental renormalization group equations
(\ref{betafns}) receive no corrections due to the dissipation. The existence of
extra dissipation terms in (\ref{ehren24}) in the Ehrenfest relation for the
momentum operator $\widehat{p}_i=-i\hbar_\mods\,\nabla_{y_i}$, which are
proportional to $\tr\,|\bar U^i|^2$, may now be compared to the generalized
Heisenberg uncertainty relations that were derived in \cite{ms1}. These extra
terms are determined by the total kinetic energy of the D-branes
and their open string excitations, and they show how the recoil of the
D-brane background produces quantum fluctuations of the classical spacetime
dynamics. The relationship with quantum uncertainty relations will be
discussed in the next section.

\section{Solutions of the Matrix D-brane Wave Equation}

In this section we shall discuss the situation
governing some of the solutions~\cite{dg7,ns} of the
class of non-linear Schr\"odinger equations
(\ref{family18}) which will be of interest to us in the context
of the quantum states of a system of many D-particles.
We may parametrize the generic wavefunctional $\Psi$
by an amplitude $\theta_1 (\vec y, t) $ and a phase $\theta_2 (\vec y, t)$ as
\be
   \Psi = {\rm exp}\left(\theta_1 + i \theta_2 \right)
\label{ampphase}
\ee
which, on account
of (\ref{family18}), satisfy the following coupled system of non-linear partial
differential equations~\cite{ns}:
\bea
0&=& \partial_t \theta_1 - 2 \nu_2 \nabla _{y_i}^2 \theta _1 - \nu_1
\nabla _{y_i}^2 \theta _2 - 4 \nu_2 \left( {\nabla _{y_i}} \theta _1 \right)^2
-2 \nu_1 {\nabla _{y_i}}\theta_1{\nabla _{y_i}} \theta _2\nn \\
0&=& \partial_t \theta_2 + 2 \mu_2 \nabla _{y_i}^2 \theta _1 + \mu_1
\nabla _{y_i}^2 \theta _2 + 4(\mu_2 + \mu_5)
\left( {\nabla _{y_i}} \theta _1 \right)^2
\nn \\
& & +\,2 (\mu_1 + \mu_4){\nabla _{y_i}}
\theta_1{\nabla _{y_i}} \theta _2 + \mu_3
\left( {\nabla _{y_i}} \theta _2 \right)^2 + \frac{1}{\hbar_{\mods}}
\,V_\mods({\vec y},\vec u;t)
\label{systemeq}
\eea
This system can be equivalently expressed in terms of the
gauge invariant set of parameters $\eta_j$ defined in (\ref{etaj21}) as
\bea
\partial_t \theta _ 1&=&2\eta _ 1 \nabla _{y_i}^2 \theta _ 1
+\frac1{2m}\,\nabla _{y_i}^2 \theta _2 +
4 \eta _1 \left({\nabla_{y_i} }\theta _1 \right)^2
+\frac1m\,{\nabla_{y_i}} \theta _ 1{\nabla_{y_i}} \theta _ 2\nn \\
\partial _t \theta _2&=&-4m\eta _ 2 \nabla _{y_i}^2 \theta _ 1
-8m\left(\eta _ 2 + \eta _5 + \eta _4 \eta_1 -(\eta_1)^2 \eta_3 \right)\left({
\nabla_{y_i}} \theta _1 \right)^2\nn \\
& &-\,2\eta_4{\nabla_{y_i}} \theta _1{\nabla_{y_i}} \theta_2
-\frac{\eta_3}{2m}\,\left({\nabla_{y_i}} \theta _ 2 \right)^2+
V_\mods(\vec y,\vec u;t)
\label{giap}
\eea
where we have selected the gauge $\mu_1=0$.
Notice that there are ambiguities in the phase function $\theta _2 $
in (\ref{ampphase}), which imply that the non-linear Schr\"odinger
equation (\ref{family18}) and the amplitude and phase equations
(\ref{systemeq}) {\it are not} completely equivalent. However, any solution
of the phase and amplitude equation yields a solution of the non-linear
Schr\"odinger equation (\ref{family18}). This is the commonly accepted point of
view \cite{ns}, and the one which we adopt in this paper.

\subsection{Mass Superselection Rules}

One of the key properties of the matrix D-brane system that we have exploited
extensively thus far is its Galilean invariance (or equivalently that the
matrix $\sigma$-model deformation defines a logarithmic conformal field
theory). It is worthwhile to first mention some basic facts concerning
Galilean invariant quantum field theories in the non-dissipative case
(${\cal D}=0$) \cite{cg}. Consider two non-relativistic
spinless fields $\phi_1$ and $\phi_2$ of masses $m_1$ and $m_2$ with a
quartic $\phi_1\phi_2$ interaction mediated by a two-body potential
$V(\vec x-\vec y)$. The corresponding Haag expansions
of their asymptotic fields are determined by Haag amplitudes $f_I(\vec
x-\vec x_a,t-t_a;\vec x-\vec x_I,t-t_I)$, where $a=1,2$ labels the two fields
and $I$ labels the different bound states of the system. Then the function
$F_I(\vec x_1-\vec x_2)$ defined through the equal time relation
\be
f_I(\vec x_1-\vec x_2;\vec x_1-\vec x_I)=\delta\left(\vec x_I-\frac{m_1\vec
x_1+m_2\vec x_2}{m_1+m_2}\right)\,F_I(\vec x_1-\vec x_2)
\label{FIdef}\ee
satisfies the usual Schr\"odinger wave equation \cite{cg}
\be
\left[-\frac12\left(\frac1{m_1}+\frac1{m_2}\right)\nabla_{x_1^i-x_2^i}^2+V(
\vec x_1-\vec x_2)\right]F_I=-\varepsilon_IF_I
\label{waveeq2body}\ee
for the stationary bound state problem. This shows that the modified Haag
amplitude
is exactly the Schr\"odinger wavefunction of the bound state, with the
reduced mass of the two particle system.

The important property used in arriving at the result (\ref{waveeq2body})
is the Galilean invariance of the system. The unitary projective
representations
of the Galilean group that arise in non-relativistic quantum mechanics cannot
be
reduced to vector representations, in contrast to the situation for most other
physically relevant groups, such as the Poincar\'e and Lorentz groups. Explicit
mass parameters appear in the phase factors of these representations
which lead to the Bargmann superselection rule \cite{bargmann} that the
sum of the masses appearing in the kinetic terms of the Hamiltonian must be
conserved in every physical
process. This implies, in particular, that the mass of
any bound state must simply be the sum of the masses of its constituents.
Note that since we are dealing with a non-relativistic theory, the energy
is not equal to the mass and indeed the energy of a
two-particle bound state differs from the sum $m_1+m_2$ by the binding
energy. Since in the case of a composite system
of $N$ heavy D-particles, the non-linear Schr\"odinger equation also stems from
the very specific Galilean invariance of the matrix D-brane dynamics, one might
expect this superselection rule to carry over to the present diffusive
situation. This would then imply that the mass appearing in the effective
kinetic term should be $Nm=N\bar g_s^{-1}/\sqrt{\alpha'}$, and the number
of colours $N$ can be absorbed into the normalization of the mass, i.e.
the $N$-body D0-brane system behaves just like a single D-brane. However,
this is not what we shall find below when we derive the
extended Haag decomposition which takes into account the diffusion terms,
i.e. of the quantum recoil of the D-particles. The massive Dirichlet string
exchange which generates contact interactions among D-branes and gives
rise to a non-linear quantum mechanical evolution equation must correctly
be incorporated in order to distinguish a single brane from the non-abelian
matrix D-branes. In the following we shall derive the modified superselection
criteria that arise in the present context.

\subsection{Stationary Solutions}

The first important class of solutions of interest to us here
are the stationary
solutions which are defined as usual~\cite{dg7} by $\partial _t {\cal P} = 0$.
This is equivalent to assuming $\partial _t \theta _ 1 =0$.
Consider the {\it free} case whereby the moduli space interaction potential
is ignored, $V_\mods=0$. Then,
setting $\theta _ 1 (\vec y, t)={\rm const.}$, one obtains from the
amplitude equation in (\ref{giap}) the plane wave solutions
$\Psi ({\vec y}, t) = \Psi _0
\,\e^{i \left( {\vec k}\cdot{\vec y} - \omega ({\vec k})t\right)}$,
with a modified dispersion relation obtained
from the resulting phase equation in (\ref{giap}):
\be
    \partial_t \theta _ 2 ({\vec y}, t)\equiv\omega ({\vec k}) =
\left(\frac{\hbar_{\mods}}{2m} + c_3 {\cal D}' \right)\vec k\,^2
\label{disp}
\ee
Notice that for the Galilean invariant but time-reversal violating
case (\ref{etainv23}),
with ${\cal D}'=0$,
the dispersion
relation (\ref{disp}) is that of a non-relativistic massive free particle.

The fact that such plane wave solutions exist is an important feature
of the present approach to the description
of the effective dynamics of a system of many D-particles.
As discussed in sections IV and V, the pertinent wave equation
of a system of $N$ D-particles, interacting via the exchange of strings,
is written down in terms of the
{\it abelianized positions} $y_i$ in (\ref{bdryfields}),
rather than an $N$-body quantum mechanics that
one would expect from the effective Yang-Mills reduction arising in the
standard low-energy description of multiple D-particle dynamics~\cite{kpdfs}.
This seems to tie in nicely with the heuristic ``fat brane'' picture
of the multiple D-brane system advocated in \cite{lms}, according
to which one treats the assembly of D-particles as a single one with string
interactions now encoded in the diffusion term
(\ref{diffcoeff11}). Had one instead started with the
Yang-Mills matrix quantum mechanics from the onset,
then one would have obtained a
{\it linear} Schr\"odinger equation with interactions taken care of by the
Yang-Mills potential. This is what had been studied in the
literature so far~\cite{kpdfs},
with no definite, general conclusions about the bound state problem
for strings and D-particles (see \cite{bound} for recent developments).
In the present approach, we effectively reduce the quantum mechanics to a
single body problem (the dynamics of the ``fat brane"~\cite{lms}),
at the cost of obtaining a
{\it non-linear} wave equation. This explains quite naturally the
 dispersion relation (\ref{disp}) in the case where string interactions
 between the D-branes are ignored. Note also that the energy is proportional
to $\bar g_s^2$ and thereby depends on the open string excitations between
the constituent D-particles.

A non-trivial issue concerns the nature of the
effective moduli space potential $V_\mods(\vec y,\vec u;t)$
among the constituent D-particles of the ``fat D-brane".
In the context of the $\sigma$-model
approach to the system of recoiling D-particles~\cite{ms1},
the effective potential $V_\mods$ among the D-particles is determined by
the Zamolodchikov $C$-function for the deformed $\sigma$-model at hand,
which according to the discussion
in \cite{ms1} and in section III, is given by the Born-Infeld
Lagrangian (\ref{NBIaction}). In the context of
the (abelianized) ``fat brane" picture,
therefore, one would expect that the relevant Born-Infeld
effective action can be reduced to the one containing only velocity
terms of the form $\tr\,\sqrt{1 - |{\bar U}^i|^2}$, given that the
coordinate-dependent commutator terms in (\ref{Fred}) may be neglected to a
first-order approximation, i.e. to leading
orders in $\sigma$-model perturbation theory \cite{ms1}.
Such a coordinate independent potential may then be shifted to zero, and hence
the free particle case $V_\mods=0$ discussed above corresponds to the
Galilean invariant multiple D-particle system in a first-order
approximation. This is similar in spirit (but not identical)
to a discussion of the ground-state properties of the
multiple D-particle system,
given that the absence of the non-commutative commutator terms
in the Born-Infeld action (\ref{NBIaction})
also guarantees unbroken supersymmetry in the standard
effective Yang-Mills target space description of multiple D-particle
systems~\cite{witten}. This commutative limit
is a saddle-point solution of the full $U(N)$ quantum gauge theory.

However, the free case is not a satisfactory
approximation if one wishes to include quantum fluctuations
about such ground states, i.e. to properly incorporate string interactions
between the D-particles which may enable them to form a bound state.
Such a case would correspond to off-shell non-conformal
backgrounds in a $\sigma$-model context which naturally appear
when one considers a quantum field theory of
D-particles~\cite{emn10,emn10b}.
Unfortunately, the precise form of the effective moduli space potential
in that case is still not known in a closed form (see \cite{ms1} for the
first few terms in a velocity expansion of the $C$-function).
For instance, it is known~\cite{kpdfs,witten} that
when a string stretches between two D-particles
one obtains a linear (confining) potential,
but the inclusion of an arbitrary number of strings
complicates the situation. Moreover, as discussed in section V, there is no
unique way of associating
the non-linear Schr\"odinger equation (\ref{family18}) to
the effective Fokker-Planck equation for the probability
distribution (\ref{FPeq12}) which was derived by world-sheet
renormalization group methods.
In particular, the precise form of the effective
potential in the linear part of the associated
Schr\"odinger equation is irrelevant for yielding the
Fokker-Planck equation (\ref{FPeq12}). This latter equation is essentially
that which governs the quantum mechanics on moduli space.
As we now show, the present approach
enables one to make some non-trivial (and generic) statements about
the existence of bound state solutions in the multiple D-particle system
under consideration, without precise knowledge of the effective
moduli space potential $V_\mods(\vec y,\vec u;t)$.

To this end we discuss a second class of stationary
solutions of the non-linear Schr\"odinger equation (\ref{family18}),
which are valid for a generic (but appropriate)
potential $V_\mods$. Such a case incorporates the complex dynamics
that binds strings and D-particles in ``fat D-brane"
configurations. Let us make the ansatz
\be
\theta_2(\vec y,t)=-m\eta_1\theta_1(\vec y)-\omega t
\label{theta2ansatz}\ee
In the Galilean invariant case (\ref{etainv23})
of relevance to us here, the function (\ref{theta2ansatz}) also solves
(\ref{giap}) and the resulting phase equation for
$\theta_1$ is
\be
2\eta_2\nabla_{y_i}^2\theta_1+4(\eta_2+\eta_5)\left(\nabla_{y_i}
\theta_1\right)^2-\frac1{2m}\,V_\mods=\frac{\omega}{2m}
\label{resphaseeq}\ee
Then the function
\be
  \varphi(\vec y)=\exp\beta\theta_1(\vec y)
\ee
where
\be
\beta=2\left(1+\frac{\eta_5}{\eta_2}\right)
  =1-\frac{{\cal D}^2}{4\alpha'\bar g_s^4}
\label{sol}\ee
with $\cal D$ the diffusion coefficient (\ref{diffcoeff11}),
satisfies the affiliated {\it linear} Schr\"odinger equation
\cite{dg7,ns}
\be
    -\frac{2m(\eta_2)^2}{\eta_2+ \eta_5}\,\nabla _{y_i}^2
\varphi+V_\mods\,\varphi
=-\omega \varphi
\label{affiliated}
\ee
Eq. (\ref{affiliated}) corresponds to an ordinary linear Schr\"odinger wave
equation
of a particle in the potential $V_\mods$, but with a shifted effective mass
\be
m^*=\beta m
\label{meff}\ee
The stationary solutions of the non-linear equation (\ref{family18})
in the case (\ref{etainv23}) are therefore given by
\be
        \Psi ({\vec y}, t ) =\varphi
({\vec y})^{\frac{4\alpha'\bar g_s^4}{4\alpha'\bar g_s^4-{\cal D}^2}}
\,\exp\left(-\frac{8i\sqrt{\alpha'}\bar g_s^3{\cal D}}
{4\alpha'\bar g_s^4-{\cal D}^2}\,\log\varphi({\vec y})^2-i\omega t \right)
\label{bound}
\ee
The wavefunctions (\ref{bound}) are square-integrable, $\Psi\in L^2(\mod)$,
if $\varphi(\vec y)^{1/\beta}$ are.

Depending on the form of the potential $V_\mods$,
such solutions may correspond to bound states
of the affiliated linear equation (\ref{affiliated}), with the same
potential $V_\mods$ but with a shifted effective mass (\ref{meff}).
It is desirable that the effective mass be positive, or otherwise
the interaction potential changes its sign. This requires $\beta>0$ in
(\ref{sol}). For a fixed string coupling constant, from (\ref{diffcoeff11}) we
see that the requirement of positivity of the effective bound state mass $m^*$
translates into a limiting velocity for the D-brane dynamics. In the abelian
case, this limiting velocity is just the speed of light \cite{bachas}, as
follows from the form of the abelian Born-Infeld action which coincides with
the standard relativistic free particle action in this case. Here we find
the bound on the velocities of the multiple D-particle system appropriate to
the non-abelian Born-Infeld action and the formation of bound state composites.
Alternatively, for a given order of magnitude of the diffusion coefficient
(\ref{diffcoeff11}) set by the recoil velocities of the D-particles, we find
that bound state solutions exist for string couplings stronger than a
critical value ${\bar g}_s^*$ given by
\be
\bar g_s^*=\sqrt{\frac{c_G}{2}\sum_{i=1}^9\tr\,|\bar U^i|^2}
\label{critical}
\ee

The solutions of the non-linear
Schr\"odinger equation differ from the bound state solutions of the affiliated
linear equation by a spatially dependent phase proportional
to the diffusion coefficient ${\cal D}$.
In the framework of the ``fat brane" picture of \cite{lms},
such bound states may be thought of as corresponding to bound state
solutions of a system of D-particles interacting via the
exchange of strings.
The existence of a critical coupling (\ref{critical}) is then
physically appealing, because it implies
the formation of bound state condensates of D-particles and strings
for strong enough string interactions, in similar spirit
to the conventional quantum field theoretic case. This is in agreement with
the fact \cite{witten} that for weak string interactions between the
constituent D-branes, the effective spacetime is commutative and no
bound states can form. When the string interactions become strong, bound
states form and render the effective target space geometry noncommutative.
Note that in the present case, the critical coupling (\ref{critical})
depends non-trivially on the (kinetic) energy scale
of the recoiling D-particles, indicating that a certain minimum amount of
energy is required to form the bound state. This velocity dependence is also
quite natural, since it means that no bound states can form if the D-particles
move too quickly relative to one another.

Let us note also that (\ref{bound}) is the analog of the Haag decomposition
in the non-diffusive case, since it is obtained from a linear Schr\"odinger
equation via a non-linear, quantum mechanical gauge transformation. This
non-linearity
implies that the bound state solutions of the affiliated linear wave equation
with shifted mass $m^*$ and potential $V_\mods(\vec y,\vec u)$ acquire
a phase that depends on $\vec y$ through
the Haag amplitude $\varphi(\vec y)$. The transition from
the non-linear equation to a related linear one is ``mediated" in part by
this $\vec y$-dependent phase. The effective bound state mass (\ref{meff})
is then the analog of the Bargmann superselection criterion in the present
case,
which we see is modified by a non-linear transformation of the original BPS
mass $m$ and also the quantum recoil diffusion.

\subsection{Gaussian Wave Solutions}

The non-linear Schr\"odinger equations
admit additional non-trivial solutions of
a Gaussian wave type. Such configurations acquire particular
importance in the ``fat D-particle" context, given that among them
there are solitary waves, and one would expect their appearance
in view of the role of D-branes as solitons in string theory.
The Gaussian wave ansatz for the wave-functional $\Psi$
is described by
\bea
     \theta _ 1 (y, t) &=& -\frac{(y-s(t))^2}{8m\sigma (t)^2} +
\frac{1}{2}\,\log\sigma(t)\nn \\
\theta_2 (y, t) &=& -\frac{1}{2m}\left(A(t)y^2 + B(t)y + C(t) \right)
\label{gauss}
\eea
where we have restricted ourselves, for simplicity,
to one spatial dimension $y$,
given that the Gaussian cases are separable~\cite{ns}. Here $s,\sigma,A,B,C$
are real-valued functions of $t$ which are determined from the amplitude
and phase equations (\ref{giap}) by equating
the different powers of $y$ that appear. This leads, in general, to a
system of coupled non-linear ordinary differential equations for these
functions
of $t$ \cite{ns,nsu}. This set of equations can be reduced to two ordinary
second
order differential equations for $\sigma$ and $s$, and $A,B,C$ may be expressed
directly in terms of the solutions of these equations.

For our purposes we note that one may explicitly
construct Gaussian wave solutions in the free case $V_\mods=0$.
Among them are also Gaussian {\it solitary} wave solutions with
time-independent width $\sigma (t) = \sigma_0 = {\rm const.}$
For the free case $V_\mods=0$, the amplitude and phase equations (\ref{giap})
imply that
a necessary condition for the existence of such solutions
is \cite{ns}
\be
\eta_2 + \eta_5 =\alpha'\bar g_s^4-\frac{{\cal D}^2}4= 0
\label{condsolgauss}
\ee
which, on account of (\ref{diffcoeff11}) and (\ref{mhbardefs}),
would imply extreme
fine tuning between the magnitude of the
velocities of the recoiling D-particles
and the string coupling $\bar g_s$. Thus it is unlikely to be met in
a generic situation, so that the class of Gaussian solitary waves do not
seem to describe the D-brane dynamics (see however the discussion below for
a different interpretation).

Nonetheless, the Gaussian ansatz (\ref{gauss}) allows one to carry out some
explicit calculations and thereby see directly the effect of the quantum
recoil through the non-linearity of the wave equation. This also enables one
to make some non-trivial consistency checks of the present formalism.
As an illustration, consider the free case
$V_\mods=0$, for which the differential equation for the function $s(t)$ is
that
of the classical motion of a free particle, $\ddot s=0$. Galilean invariance
may
then be used to set $s(t)=0$ for all $t$, because a general solution $s(t)=
u_0t+s_0$ can be obtained from the $s=0$ one by a $G(1)$ transformation. The
remaining differential equations can be easily integrated by quadratures to
give
the solutions \cite{nsu}
\bea
\sigma(t)&=&\sigma_0\,\sqrt{\left(\frac{2\bar g_s^2t}{\sqrt{\alpha'}\sigma_0^2}
+f_0\right)^2
+\beta}\nn\\A(t)&=&\frac{16}{(\alpha')^{3/2}\bar
g^3_s\sigma_0^2}\,\frac{\frac{2\bar
g_s^2t}{\sqrt{\alpha'}\sigma_0^2}+f_0-\frac{{\cal D}}{2\sqrt{\alpha'}\bar
g_s^2}}{\sqrt{\left(\frac{2\bar g_s^2t}{\sqrt{\alpha'}\sigma_0^2}+f_0
\right)^2+\beta}}\nn\\B(t)&=&0\nn\\C(t)&=&-\frac4{\sqrt{\alpha'\beta}}
\,\arctan\left(\frac{\frac{2\bar g_s^2t}{\sqrt{\alpha'}\sigma_0^2}+f_0}
{\sqrt\beta}\right)+C_0
\label{completefreesoln}\eea
where $\sigma_0$, $f_0$ and $C_0$ are appropriate integration constants and
$\beta$ is defined in (\ref{sol}).
The complete solution for the wavefunction $\Psi$ is then obtained by
substituting
(\ref{completefreesoln}) into (\ref{gauss}) and (\ref{ampphase}) with $s=0$.
{}From this solution it is straightforward to compute expectation values of
operators in terms of Gaussian integrals, and in particular
arrive at the variance relation
\cite{nsu}
\be
\Delta y\,\Delta p=2\bar g_s\,\sqrt{1+\left(\frac{2\bar g_s^2t}{\sqrt{\alpha'}
\sigma_0^2}+f_0-\frac{{\cal D}}{2\sqrt{\alpha'}\bar g_s^2}\right)^2}
\label{varreln}\ee

The uncertainty relation (\ref{varreln})
corresponds to that of a minimum uncertainty wavepacket,
with effective Planck constant determined by the string coupling, as is
conjectured in \cite{ms1}. In fact, eq. (\ref{varreln}) has a remarkably
similar form to the modified Heisenberg uncertainty relations that were
derived in \cite{ms1} directly from the
string $\sigma$-model genus expansion, in that it contains a diffusion term
proportional to the total kinetic energy of the D-particles and their open
string excitations which implies that the accuracy with which one can
measure position and momentum in a system of D-particles depends on the
(recoil) energy that arises in the measurement process.
Such an energy dependence is characteristic of string-modified uncertainties
and quantum gravitational effects. At time $t=0$ we can expand the
right-hand side of (\ref{varreln}) for slowly moving branes to get
\be
\Delta y\,\Delta p\simeq\bar g_s\left(2+f^2_0
\right)-\frac{f_0c_G}{\bar g_s}
\sum_{i=1}^9\tr\,|\bar U^i|^2
\label{varrelnslow}\ee
To leading orders in $\sigma$-model
perturbation theory and for weakly coupled strings, the momentum fluctuations
$\Delta p$ and recoil velocities
$\bar U$ coincide up to the BPS mass of the D-branes \cite{ms1}. Choosing
${\rm sgn}\,f_0=-1$, it follows that
(\ref{varrelnslow}) has the usual form of the string-modified Heisenberg
uncertainty relation:
\be
\Delta y\,\Delta p\simeq\bar g_s\left(2+f_0^2\right)+\frac{N|f_0|c_G
\alpha'}{\bar g_s^3}\,(\Delta p)^2
\label{dydp}\ee
The solution to the equation for the
extrema of $\Delta y$ in (\ref{dydp}) as a function of $\Delta p$
leads to the usual minimum measurable length,
and hence to the standard expectations that it is not
possible to probe distances smaller that
the intrinsic string length or the spacetime Planck
length using only D-particle probes (Note that for $f_0\geq0$ there is no bound
on the measurability of lengths in the spacetime).
Notice that one may in fact use the complete form of the phase space
uncertainty relations derived in \cite{ms1}, along with the modified
Ehrenfest relations (\ref{ehren24}),
as a necessary criterion to determine in general if a
candidate wavefunction represents the true quantum states of the D-particles
and strings. We shall not do so here, but rather carry on with the heuristic
analysis and present one final possible description of
the quantum dynamics of the system.

\subsection{Solitary Wave Solutions}

There exist other types of solitary waves which may occur for
certain regions of the two-dimensional parameter space
spanned by $\eta_2$ and $\eta_5$.
The construction of such solutions in the free case $V_\mods=0$ may be
carried out as follows \cite{ns}.
For the Galilean invariant case under consideration,
if $\theta_1$ and $\theta_2$ are solutions of the equations
(\ref{theta2ansatz}) and (\ref{resphaseeq}), then
the Galilean boosted amplitude and phase
\bea
{\tilde \theta_1} ({\vec y}, t) &=& \theta_1 ({\vec y} - {\vec u} t)\nn \\
{\tilde \theta_2} ({\vec y}, t) &=& \theta_2 ({\vec y}- {\vec u} t, t) -
m\vec u\cdot\vec y-\frac{m\vec u\,^2}2\,\eta_3 t
\label{boost}
\eea
with ${\vec u}$ a constant velocity,
are {\it non-stationary} solutions of (\ref{giap}).
Such transformations take the original Galilean invariant solution
of the non-linear Schr\"odinger equation (\ref{family18}),(\ref{etainv23}),
with $\eta_4=0,\eta_3=-1$,
to solutions in the more general case
\be
-2\eta_1(1 + \eta_3) + \eta_4=0
\label{moregencase}\ee
The resulting solution for the wave-functional $\Psi$ then
reads
\bea
\Psi ({\vec y}, t) = \Psi _0\,\exp\left(\theta_1 ({\vec y} - {\vec u} t)
-i\left[4m\eta_1\theta_1 ({\vec y}-{\vec u}t)+
m{\vec u}\cdot{\vec y}+\frac{m\vec u\,^2}2\,\eta_3t+\omega t \right]\right)
\label{ngs}
\eea
where $\Psi_0$ is a constant. The associated probability
density ${\cal P}=\Psi^\dagger\Psi$
is itself a {\it solitary wave}, moving with constant speed
and without changing its shape:
\be
     {\cal P}({\vec y}, t ) = \left|\Psi _0 \right|^2\,
\exp2\theta_1 ({\vec y}-{\vec u} t )
\label{probgs}
\ee
In the particular case where $\eta_2 \left(\eta_2 + \eta_5 \right) < 0$,
i.e. $\beta<0$, one obtains
square-integrable solitary wave solutions of the form~\cite{ns}
\bea
\Psi ({\vec y}, t)&=&\Psi _0\,\cosh\left( {\vec k}\cdot({\vec y}
- {\vec u} t ) \right)^{\frac{4\alpha'\bar g_s^4}{4\alpha'\bar g_s^4
-{\cal D}^2}}\nn \\
& &\times\,\exp\left[-\frac{8i\sqrt{\alpha'}\bar g_s^5{\cal D}}{4\alpha'
\bar g_s^4-{\cal D}^2}\,\log\cosh\left({\vec k}\cdot({\vec y}-{\vec
u}t)\right)-
\frac{16\alpha'\bar g_s^8}{4\alpha'\bar g_s^4-
{\cal D}^2}\,\vec k\,^2 t\right]
\label{wave}
\eea
Notice that the condition $\beta<0$ is opposite to that
for the existence of positive effective mass bound states
in the affiliated linear Schr\"odinger equation (\ref{affiliated}).
This implies that such coherent solitonic states
cease to exist for string couplings stronger
than the critical value (\ref{critical}),
for given recoil velocities of the D-particles.

Such solitary waves may play an important role in the underlying dynamics
of a system of D-particles. The fact that they arise in the
``fat brane" picture indicates that the latter degree of freedom
constitutes an effective
coarse-grained description of macroscopic quantum coherent states
that may characterize (under the specified conditions) a certain
phase of
the ground state of a system of D-particles
interacting among themselves via the exchange of strings.
{}From this point of view, the critical coupling situation
(\ref{condsolgauss}) where ${\bar g}_s = {\bar g}_s^*$,
corresponding to Gaussian solitary waves,
may be thought of as a critical regime
separating the bound state phase from the
solitary wave coherent phase. It should be noted, however, that although
the wavefunctions (\ref{bound}) and (\ref{wave}) are non-analytic around
$\bar g_s=\bar g_s^*$, this change of solution should not be regarded as
some phase transition in string coupling constant space. Rather, when the
string interactions become weak enough to untighten the bound states of
D-particles and strings, the quantum dynamics is described by minimum
uncertainty
wavepackets which produce the appropriate smearing to the quantum
spacetime. Outside of this regime the D-particles behave as freely propagating
solitary waves lending another interpretation to the notion of D-branes
as string theoretic solitons.
It should also be stressed that this heuristic picture is complementary
to the picture discussed in
\cite{ms1}, where the microscopic structure of
groups of recoiling D-particles under the exchange of strings
has been studied in detail in a $\sigma$-model context.
In the present picture one finds an effective
quantum mechanics for a ``coarsed grained" description of
the multiple D-particle system, in which ``lumps" of D-particles
are described as single bodies obeying a
{\it non-linear} diffusive Schr\"odinger dynamics which encodes their
interactions. As mentioned before, this analysis is at present preliminary
and must thought of as a heuristic picture for the D-brane dynamics, given
that it neither incorporates supersymmetry nor the precise form of the
effective moduli space potential $V_\mods$. The above discussion has only
centered around the free case as an illustration, which is correct only to
a certain approximation, since the abelianization of the collective coordinates
of the D-particles, on which the ``free" particle picture is based, is
not accurate beyond this mean field approximation.

\section{Conclusions}

The above discussion completes the first analysis concerning the construction
of a Schr\"odinger-type wave equation (here non-linear) for the wavefunctional
of a collection of D-particles. This equation is the present proposal for the
complicated quantum dynamics of Dirichlet branes, at least within the framework
of worldsheet $\sigma$-model perturbation theory. Further analysis should
concentrate on the applicability of such studies to interferometric devices,
which might probe Planckian physics. The non-linear Schr\"odinger dynamics that
we have described above to characterize generic systems of D-branes should also
be compared to the linear modifications of Schr\"odinger wave equations in open
quantum mechanical systems interacting with a stochastic type environment. Such
types of interactions are customarily assumed in many approaches to quantum
gravity.

There are, in general, arguments against non-linear Schr\"odinger wave
equations, mostly stemming from the fact that non-linearities may lead to
superluminal propagation, i.e. motion which goes faster than the speed of light
(see the discussion in \cite{dg8} and references therein). In the D-brane
picture advocated above, there is the advantage of expressing the system in a
``closed'' but non-linear form, i.e. writing a self-consistent quantum equation
which comprises only known degrees of freedom, including recoil. In the above
analysis we did not use environment operators, which are usually unknown and
assumed to be generically of a stochastic form. In the present approach we have
demonstrated the existence of quantum diffusion and of
stochasticity~\cite{ms1,emn2}
in the sense of Gaussian probability distributions obtained explicitly as a
result of summing over all worldsheet genera (i.e. coming from the quantum
string theory).

Moreover, as shown in \cite{ms1}, the resulting off-shell target space action
underlying the Schr\"odinger dynamics near a fixed point in moduli space (where
the entire approach is valid) is the non-abelian Born-Infeld action, which
implies a limiting velocity (equal to the speed of light) for the D-particle
dynamics \cite{bachas}. This feature prevents superluminal propagation. Thus
the issues of superluminal propagation due to a non-linear Schr\"odinger wave
equation, which were originally advocated in \cite{gisin}, cannot apply to the
D-particle cases. These issues express such conceptual difficulties as the
possibility of using long-ranged quantum correlations to send instantaneous
signals, and of communicating among different quantum mechanical worlds in a
``multiple world'' interpretation.
As noted in \cite{dg8}, since non-linear gauge
transformations on the wavefunctional can be found which linearize the
non-linear Schr\"odinger wave equation, such claims cannot be correct in
general. The Fokker-Planck equation that we have derived for the probability
density may correspond to a linear Schr\"odinger wave equation, but of an {\it
open} system, i.e. including
interactions with environmental degrees of freedom (in agreement with the
analysis of \cite{emn10,gisin}). However, the non-linear wave equation
involving only functionals of $\Psi$, and thereby closing the system, is
preferred. We do not need to appeal here to the unknown environmental degrees
of freedom to describe the recoil-induced diffusion of D-brane dynamics.

Although the arguments for superluminal propagation are overcome by the
existence of a Born-Infeld Lagrangian for the D0-brane dynamics, a more direct
approach is desired since the Fokker-Planck diffusion equation implies that an
external environment is present for the D-brane dynamics. This may be argued
through the fact that the non-linear effects that we have considered cannot
produce superluminal propagation because the present formalism is
self-consistent only for weakly-coupled strings with $\bar g_s\to0$
(corresponding to heavy D-particles and hence very small velocities). Light
string states are then obtained by $S$-duality under which the resulting
Schr\"odinger wave equation may indeed become linear.

Notice also that the time-reversal violation of the effective wave equation
that we have obtained is implied directly (through entropy production) by the
fact that decoherence is present in the system of D-particles. However, if
$CPT$ is conserved in a standard string theory, then $T$ violation should occur
if $CP$ symmetry breaking is present. In the Born-Infeld Lagrangian one may
have induced $CP$ violation in four dimensions, for example, by the
introduction of topological instanton terms. In string phenomenology such terms
arise quite naturally upon compactification of the target space. The present
approach which allows decoherence (due to the energy dependence of the quantum
uncertainties) therefore provides a natural explanation of the supression of
$CP$ symmetry in string theory. But generally in decohering quantum mechanical
systems $CPT$ is not conserved, so the above approach also differentiates
between standard particles and recoiling D-particles.

We close by mentioning that non-linear quantum dynamics
may be desirable from another point of view discussed
recently in \cite{czachor}. The non-linear evolution
of correlated systems (i.e. those involving the dynamics of
many ``particle'' states) may be free from
the contamination by negative probabilities which inflict
the corresponding linear case. The
D-brane systems we are considering here
always involve many D-branes interacting with each other
via the exchange of closed string states. In the present picture~\cite{ms1}
single D-particles are only viewed as a limiting case
of a multi-brane system where the constituent
branes are separated at macroscopic distances compared to the
string length $\sqrt{\alpha'}$. The non-linear dynamics, therefore, which
we proposed in this article, may be essential for guaranteeing
complete positivity of the system
in the generalized sense discussed in \cite{czachor}.
If one accepts the modern viewpoint of string-inspired phenomenology
of the observable world~\cite{extra},
according to which the latter is represented as a D-brane
interacting with other branes via the exchange
of closed string bulk states only, then
the above-mentioned issues may be of crucial importance
for a consistent formulation.

\acknowledgments{N.E.M. wishes to thank J. Ellis and
D. Nanopoulos for discussions.
R.J.S. would like to thank G. Semenoff for hospitality at
the University of British Columbia, where part of this work was carried out.
R.J.S. would also like to thank the organisors and participants of the
PIms/APCTP/CRM workshop ``Particles, Fields and Strings '99'', which was held
at the Pacific Institute for the Mathematical Sciences, University of
British Columbia, August 2--20 1999, for having provided a stimulating
environment for research.
The work of N.E.M. was supported in part by a PPARC (U.K.) Advanced
Fellowship. The work of R.J.S. was supported in part by the Danish
Natural Science Research Council.}

\section*{Appendix} 

In this Appendix we present a world-sheet formalism 
in support of the 
identification of the world-sheet partition function 
of a Liouville string with the 
probability density in moduli space rather than a wavefunctional, as
happens in the case of critical strings.

\begin{figure}[htb]
\epsfig{figure=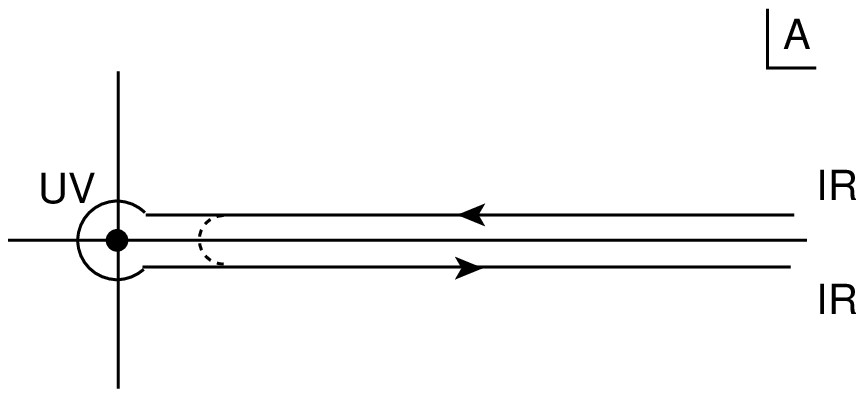}
\caption{Contour
of integration in the analytically-continued
(regularized) version of $\Gamma (-s)$ for $ s \in Z^+$.
The quantity A denotes the (complex) world-sheet area. 
This is known in the literature as the Saalschutz contour,
and has been used in
conventional quantum field theory to relate dimensional
regularization to the Bogoliubov-Parasiuk-Hepp-Zimmermann
renormalization method. Upon the interpetation of the 
Liouville field with target time, this curve
resembles closed-time-paths in non-equilibrium field theories.} 
\label{fig1}
\end{figure}

We commence our analysis 
by considering the correlation functions among vertex operators 
in a generic Liouville theory, viewing the Liouville field
as a local renormalization-group scale on the world sheet~\cite{emn}.
Standard computations\cite{goulian} yield for an $N$-point correlation
function among world-sheet integrated
vertex operators $V_i\equiv \int d^2z V_i (z,{\bar z}) $ :
\be
A_N \equiv <V_{i_1} \dots V_{i_N} >_\mu = \Gamma (-s) \mu ^s
<(\int d^2z \sqrt{{\hat \gamma }}e^{\alpha \phi })^s {\tilde
V}_{i_1} \dots {\tilde V}_{i_N} >_{\mu =0}
\label{C12}
\ee
where the tilde denotes removal of the
Liouville  field $\phi $ zero mode, which has been
path-integrated out in (\ref{C12}).
The world-sheet scale $\mu$ is associated with cosmological
constant terms on the world sheet, which are characteristic
of the Liouville theory.
The quantity $s$ is the sum of the Liouville anomalous dimensions
of the operators $V_i$
\be
s=-\sum _{i=1}^{N} \frac{\alpha _i}{\alpha } - \frac{Q}{\alpha}
\qquad ; \qquad \alpha = -\frac{Q}{2} + \frac{1}{2}\sqrt{Q^2 + 8}
\label{C13}
\ee
The $\Gamma $ function can be regularized\cite{kogan,emn}
(for negative-integer
values of its argument) by
analytic coninuation to the complex-area plane using the
the Saaschultz contour
of Fig. \ref{fig1}. Incidentally, this yields the possibility
of an increase of the running central charge
due to the induced oscillations of the dynamical
world sheet area (related to the Liouville zero mode).
This is associated with an oscillatory solution
for the Liouville central charge near the fixed point.
On the other hand, the bounce intepretation
of the infrared fixed points of the flow,
given in refs. \cite{kogan,emn},
provides an alternative picture
of the overall monotonic change
at a global level in target space-time.

To see technically why the above formalism 
leads to a breakdown in the 
interpretation of the correlator $A_N$ 
as a target-space string amplitude, which in turn leads to  
the interpretation of the world-sheet
partition function as a probability density rather than a 
wave-function in target space, 
one first expands the Liouville
field in (normalized) eigenfunctions  $\{ \phi _n \}$
of the Laplacian $\Delta $ on the world sheet
\be
 \phi (z, {\bar z}) = \sum _{n} c_n \phi _n  = c_0 \phi _0
 + \sum _{n \ne 0} \phi _n \qquad \phi _0 \propto A^{-\frac{1}{2}}
\label{C14}
\ee
with $A$ the world-sheet area,
and
\be
   \Delta \phi _n = -\epsilon_n \phi _n  \qquad n=0, 1,2, \dots,
\qquad \epsilon _0 =0
\qquad (\phi _n, \phi _m ) = \delta _{nm}
\label{C15}
\ee
The result for the correlation functions (without the Liouville
zero mode) appearing on the right-hand-side of eq. (\ref{C12})
is, then
\bea
{\tilde A}_N \propto &\int & \Pi _{n\ne0}dc_n exp(-\frac{1}{8\pi}
\sum _{n\ne 0} \epsilon _n c_n^2 - \frac{Q}{8\pi}
\sum _{n\ne 0} R_n c_n + \nn \\
~&~&\sum _{n\ne 0}\alpha _i \phi _n (z_i) c_n )(\int d^2\xi
\sqrt{{\hat \gamma }}e^{\alpha\sum _{n\ne 0}\phi _n c_n } )^s
\label{C16}
\eea
with $R_n = \int d^2\xi R^{(2)}(\xi )\phi _n $. We can compute
(\ref{C16}) if we analytically continue \cite{goulian}
$s$ to a positive integer $s \rightarrow n \in {\bf Z}^{+} $.
Denoting
\be
f(x,y) \equiv  \sum _{n,m~\ne 0} \frac{\phi _n (x) \phi _m (y)}
{\epsilon _n}
\label{fxy}
\ee
one observes that, as a result
of the lack of the zero mode,
\be
   \Delta f (x,y) = -4\pi \delta ^{(2)} (x,y) - \frac{1}{A}
\label{C17}
\ee
We may choose
the gauge condition  $\int d^2 \xi \sqrt{{\hat \gamma}}
{\tilde \phi }=0 $. This determines the conformal
properties of the function $f$ as well as its
`renormalized' local limit
\be
   f_R (x,x)=lim_{x\rightarrow y } (f(x,y) + {\rm ln}d^2(x,y))
\label{C18}
\ee
where  $d^2(x,y)$ is the geodesic distance on the world sheet.
Integrating over $c_n$ one obtains
\bea
~&& {\tilde A}_{n + N} \propto
exp[\frac{1}{2} \sum _{i,j} \alpha _i \alpha _j
f(z_i,z_j) + \nn  \\
~&&\frac{Q^2}{128\pi^2}
\int \int  R(x)R(y)f(x,y) - \sum _{i} \frac{Q}{8\pi}
\alpha _i \int \sqrt{{\hat \gamma}} R(x) f(x,z_i) ]
\label{C19}
\eea

We now consider
infinitesimal Weyl shifts of the world-sheet metric,
$\gamma (x,y) \rightarrow \gamma (x,y) ( 1 - \sigma (x, y))$,
with $x,y$ denoting world-sheet coordinates.
Under these,
the correlator $A_N$
transforms as follows~\cite{emn} 
\bea
&~&
\delta {\tilde A}_N \propto
[\sum _i h_i \sigma (z_i ) + \frac{Q^2}{16 \pi }
\int d^2x \sqrt{{\hat \gamma }} {\hat R} \sigma (x) +    \nn \\
&~&
\frac{1}{{\hat A}} \{
Qs \int d^2x \sqrt{{\hat \gamma }} \sigma (x)
       +
(s)^2 \int d^2x \sqrt{{\hat \gamma }} \sigma (x) {\hat f}_R (x,x)
+  \nn \\
&~&
Qs \int \int d^2x d^2y
\sqrt{{\hat \gamma }} R (x) \sigma (y) {\hat {\cal
 G}} (x,y) -
  s \sum _i \alpha _i
  \int d^2x
  \sqrt{{\hat \gamma }} \sigma (x) {\hat {\cal
 G}} (x, z_i) -   \nn \\
&~&
 \frac{1}{2} s \sum _i \alpha _i{\hat f}_R (z_i, z_i )
  \int d^2x \sqrt{{\hat \gamma }} \sigma (x)
-    \nn \\
&~&
 \frac{Qs}{16\pi} \int
  \int d^2x d^2y \sqrt{{\hat \gamma (x)}{\hat \gamma }(y)}
  {\hat R}(x) {\hat f}_R (x,x) \sigma (y)\} ] {\tilde A }_N
\label{dollar}
\eea
where the hat notation denotes transformed quantities,
and
the function  ${\cal G}$(x,y)
is defined as
\be
  {\cal G}(z,\omega ) \equiv
f(z,\omega ) -\frac{1}{2} (f_R (z,z) + f_R (\omega, \omega ) )
\label{C20}
\ee
and transforms simply under Weyl shifts~\cite{emn}.
We observe from (\ref{dollar}) that
if the sum of the anomalous dimensions
$s \ne 0$ (`off-shell' effect of
non-critical strings), then there are
non-covariant terms in
(\ref{dollar}), inversely proportional to the
finite-size world-sheet area $A$.
Thus the generic correlation 
function $A_N$ does not have a well-defined 
limit as $A \rightarrow 0$. 

In our approach to string time we identify~\cite{emn} 
the target time as $t=\phi_0=-{\rm log}A$, 
where $\phi_0$ is the world-sheet zero mode of the Liouville field.
The normalization follows from
a consequence 
of the canonical form of the kinetic term for the Liouville field $\phi$ 
in the Liouville $\sigma$ model~\cite{aben,emn}. 
The opposite flow of the target time, as compared to that of the 
Liouville mode, is, on the other hand, a consequence
of the `bounce' picture~\cite{kogan,emn} for Liouville flow of Fig. 1.
In view of this, the above-mentioned induced time (world-sheet scale 
$A$-) dependence
of the correlation functions $A_N$ implies the
breakdown of their interpretation as
well-defined $S$-matrix elements,
whenever there is a departure from criticality $s \ne 0$. 

In general, this is a feature of non-critical strings
wherever the Liouville mode is viewed as a local 
renormalization-group scale
of the world sheet~\cite{emn}. In such a case,
the central charge of the theory
flows continuously with the world-sheet scale $A$,
as a result of the Zamolodchikov
$c$-theorem \cite{zam}. In contrast, the screening operators
in conventional strings 
yield quantized values\cite{aben}.
Due to the analytic continuation curve illustrated in Fig. \ref{fig1},
we observe that upon intepreting the Liouville field $\phi$ 
as time~\cite{emn}: $t \propto {\rm log}A$, 
the contour of Fig. 1 represents evolution 
in both directions of time between fixed points of the 
renormalization group: $ {\rm Infrared} ~ {\rm fixed} ~ 
{\rm point}  \rightarrow {\rm  Ultraviolet} ~ {\rm fixed}
~{\rm point} \rightarrow
 {\rm Infrared} ~ {\rm fixed} ~ {\rm point}$.

When one integrates over the Saalschultz contour in fig. \ref{fig1}, 
the integration
around the simple pole at $A=0$ yields an imaginary part~\cite{kogan,emn},
associated with the instability of the Liouville vacuum. We note, on the  
other hand, that the integral around the dashed contour
shown in Fig. \ref{fig1}, 
which does not encircle the pole at $A=0$, is well defined.
This can be intepreted as a well-defined $\nd{S}$-matrix element,
which is not, however, factorisable into a product of 
$S-$ and $S^\dagger -$matrix elements, due to the 
$t$ dependence acquired after the identification 
$t=-{\rm log}A$. 

Note that 
this formalism is similar to the 
Closed-Time-Path (CTP) formalism used in non-equilibrium 
quantum field theories~\cite{ctp}.
Such formalisms are characterized by a `doubling of degreees of 
freedom' (c.f. the two directions of the time (Liouville scale) 
curve of figure \ref{fig1}, 
in each of which one 
can define a set of dynamical fields in target space). 
As we discussed above, this prompts one 
to identify the corresponding 
Liouville correlators $A_N$ with $\nd{S}$-matrix elements 
rather than $S$-matrix elements in target space. 
Such elements act on the 
density matrices $\rho={\rm Tr}_{{\cal M}}|\Psi><\Psi|$ 
rather than wave vectors $|\Psi>$ in target space of the string:
$\rho_{out} = \nd{S} \rho_{in}$ (c.f. the analogy 
with the $S$-matrix, $|out> =S|in>$). 

This in turn implies that the world-sheet partition function ${\tilde 
{\cal 
Z}}_{\chi,L}$ 
of a Liouville string at a given world-sheet genus $\chi$,
which is connected to the generating functional 
of the Liouville correlators $A_N$, 
when 
{\it defined} over the closed Liouville (time) path (CTP) 
of figure \ref{fig1},  
can be associated 
with the {\it probability density} (diagonal element of a density matrix)
rather than the 
wavefunction in the space of couplings. Indeed, one has
\begin{equation}
{\tilde {\cal Z}}_{\chi,L}[g^I] = 
\int_{CTP} d\phi_0 {\cal Z}_{\chi,L}[\phi_0, g^I] 
\label{liouvpartfnct}
\end{equation}
where 
$\{ g^I \}$ denotes the set of couplings of the (non-conformal) deformations, 
$\phi_0 \sim {\rm ln}A$ is the Liouville zero mode, and 
$A$  is the world-sheet  area (renormalization-group scale). 
If one naively interprets 
${\cal Z}_{\chi,L}[\phi_0,g^I]$ as a wavefunctional 
in moduli space $\{ g^I \}$, $\Psi [\phi_0, g^I ]$, 
then,  
in view of the double contour of figure \ref{fig1},
over which ${\tilde {\cal Z}}_{\chi,L}$ is defined, 
one 
encounters at each slice of constant $\phi_0$ 
a product of $\Psi [\phi_0, g^I]\Psi^\dagger [\phi_0, g^I]$, 
the complex conjugate wavefunctional corresponding to the 
second branch of the contour of opposite sense to the branch 
defining $\Psi [\phi_0,g^I]$. This is analogous to the 
doubling of degrees of freedom in conventional 
thermal field theories~\cite{ctp}. 
Such products represent clearly 
probability densities ${\cal P}[t,g^I]$ 
in moduli space of the non-critical strings
upon the identification of 
the Liouville zero mode $\phi_0$ with the target time $t$~\cite{emn}. 

In the above spirit, one may then consider 
the (formal) summation over world-sheet topologies $\chi$, and 
identify the summed-up world-sheet partition function 
$\sum_{\chi} {\cal Z}_{\chi,L}[\phi_0, g^I]$ 
with the associated probability density in moduli space. 
In the case of D-particles, discussed in this work, 
the moduli space coincides with the configuration space (collective) 
coordinates of the D-particle 
soliton, and hence the corresponding probability density 
is associated with the position of the D-particle in target space. 
We stress once again that the above conclusion 
is based on the 
crucial assumption of the 
definition of the Liouville-string world-sheet partition 
function over the closed-time-path of figure \ref{fig1}. 
As we demonstrate in the main text, 
the specific D--brane example provide us  
with highly non-trivial 
consistency checks of this approach.

Before closing this appendix we would like to give an explicit demonstration
of the above ideas for the specific (simplified) case  
of recoiling (Abelian) D-particles. We shall demonstrate below
that, upon considering the non-critical $\sigma$-model of a recoiling 
D-particle at a fixed world-sheet (Liouville) scale $\phi_0={\rm ln}A$, 
and identifying the 
Liouville mode with the target time, 
the Euclideanized 
world-sheet partition function can describe a  probability density
in moduli (collective coordinate) space. 

To this end, let us first consider the pertinent 
$\sigma$ model partition 
function for a D-particle, at tree level and 
in a {\it Minkowskian} world-sheet $\Sigma$ formalism:
\begin{equation} 
{\cal Z}_{\chi=0, L} = \int (DX^i)~ e^{-i\frac{1}{4\pi \alpha'} \int _{\Sigma} \partial X^i 
{\overline \partial} X^j \eta_{ij} - i\frac{1}{2\pi \alpha'} 
\int _{\partial \Sigma} \left(\epsilon g_i^C + 
g_i^D \frac{1}{\epsilon} \right) \partial_n X^i }
\label{dparticlepf}
\end{equation} 
where $\epsilon^{-2} \sim {\rm ln}\Lambda ^2={\rm ln}A$ 
(c.f. (\ref{epLambdaid})), 
on account of the 
logarithmic algebra~\cite{kmw12}. In our approach
$\epsilon^{-2}$ is identified with the 
target time. 
This is why in (\ref{dparticlepf}) we have not path-integrated
over $X^0$, but we consider 
an integral only over the spatial collective coordinates 
$X^i, i=1, \dots 9$ of the D-particle. The combination of $\sigma$-model 
couplings 
$ \epsilon g_i^C + g_i^D \frac{1}{\epsilon} $ may be identified with the
generalized (Abelian) position
$\epsilon Y^i$ of the recoiling D-particle (\ref{recoilY}).
Notice that, since here we have already identified the time with the 
scale $\epsilon^{-2} >0$, the step function in the recoil deformations
of the $\sigma$-model (\ref{recoilops}) 
acquires trivial meaning.
We shall come back to a discussion on how one can incorporate
a world-sheet dependence in the time coordinate later on.  

Suppose now that, following the spirit of critical strings~\cite{hlp6},
one identifies the Minkowskian world-sheet partition function 
(\ref{dparticlepf})  
with a wavefunctional $\Psi[Y^i, \phi_0=t]$.
The probability density in $Y^i$ space, 
${\cal P}[Y^i, t]=\Psi[Y^i,t]\Psi^*[Y^i,t]$,  
reads in this case:
\begin{eqnarray} 
&~& \left|{\cal Z}_{\chi=0, L}[Y^i,t]\right|^2 = 
\int DX^i \int DX'^j e^{-i\frac{1}{4\pi \alpha'} 
\int _{\Sigma} \partial X^i 
{\overline \partial} X^j \eta_{ij} + i\frac{1}{4\pi \alpha'}
\int _{\Sigma} \partial X'^i 
{\overline \partial} X'^j \eta_{ij} 
-i \frac{1}{2\pi \alpha'} 
\int _{\partial \Sigma} \epsilon Y_i(t)\partial_n (X^i - X'^i)} =
\nonumber \\
&~& \left(\int DX_{-}^i e^{i\frac{1}{4\pi\alpha'}\int _{\Sigma} \partial X_{-}^i 
{\overline \partial} X_{-}^j \eta_{ij} 
-i \frac{1}{2\pi \alpha'}\int _{\partial \Sigma} Y_i(t) \partial_n X_{-}^i }\right) \otimes 
(\left(\int DX_{+}^i e^{-i\frac{1}{4\pi \alpha'} \int _{\Sigma} 
\partial X_{+}^i 
{\overline \partial} X_{+}^j \eta_{ij}}\right)~,
\label{factorization}
\end{eqnarray}
where $X_{\pm}^i=X^i \pm X'^i $. 
Upon passing to a Euclidean world-sheet formalism, 
and taking into account that the $Y_i$ independent factor 
can be absorbed in appropriate normalization of the 
$\sigma$-model correlators, one then 
proves our statement
that $\sigma$-model partition functions in non-critical strings
can be identified with moduli space probability densities. 

Notice that similar conclusions can be reached even in the case
where the time $X^0$ is included in the analysis as a full fledged 
world-sheet field and {\it is only eventually} 
identified with the Liouville mode. 
In such a case, by considering the probability density as above, 
one is confronted with path integration over $X_{\pm}^0 = X_0 \pm X'^0$ 
$\sigma$-model fields, which also appear in the arguments
of the step function operators $\Theta_\epsilon ( X_{\pm}^0 )$
in the recoil deformations (\ref{recoilops}),  
that 
are non trivial in this case. 
However, upon Liouville dressing and the {\it requirement} 
that the  Liouville mode be identified with the target  time,
one is forced to restrict oneself on the hypersurface $X_{-}=0$
in the corresponding  path integral $\int DX_{+}^0 DX_{-}^0 ( \dots )$. 
As a consequence, one is then left
with a world-sheet partition 
function 
integrated only over 
the Liouville mode $X_{+}=2\phi$ (c.f. ${\tilde {\cal Z}}$ 
in (\ref{liouvpartfnct})), 
and hence the identification of 
a Liouville string partition 
function with a probability density in 
moduli space  
is still valid, upon passing onto a Euclideanized
world-sheet formalism. It can also be seen, in a straightforward manner,
that summing upon higher world-sheet topologies, 
as in \cite{ms1}, will not change this conclusion.

\vfill
\pagebreak

\newpage

\end{document}